\documentclass[a4paper, 11pt]{article}

\usepackage{graphicx}
\usepackage{epsf,amsmath,bbold,amsfonts,stmaryrd}
\usepackage{multirow}
\usepackage{makecell} 
\usepackage[utf8]{inputenc}
\usepackage{mathrsfs}
\usepackage{appendix}
\usepackage{amssymb}
\usepackage{float}
\usepackage{color}
\usepackage{cite}
\usepackage{hyperref}
\hypersetup{pageanchor=false}
\usepackage{indentfirst}
\usepackage{url}
\usepackage{float}
\usepackage{caption}
\usepackage[numbers,square,comma,sort&compress,merge]{natbib}

\usepackage{subfigure}

\usepackage{ulem}

\def\be{\begin{equation}}
\def\ee{\end{equation}}

\def\bea{\begin{eqnarray}}
\def\eea{\end{eqnarray}}

\def\ba{\begin{array}}
\def\ea{\end{array}}

\def\bc{\begin{center}}
\def\ec{\end{center}}

\def\bl{\begin{flushleft}}
\def\el{\end{flushleft}}

\def\br{\begin{flushright}}
\def\er{\end{flushright}}

\def\bi{\begin{itemize}}
\def\ei{\end{itemize}}

\def\bt{\begin{tabular}}
\def\et{\end{tabular}}

\numberwithin{equation}{section}

\begin{document}

\title{Bayesian Analysis of Massive Boson Star Models for Sagittarius A*
	Using Near-Infrared Astrometry Data}

\author{Xiangyu Wang$^{1}$, Tian-chi Ma$^{1}$, Minyong Guo$^{2, 3\ast}$, Hai-Qing Zhang$^{1, 4\dagger}$
}
\date{}

\maketitle

\vspace{-10mm}

\begin{center}
{\it
	$^1$ Center for Gravitational Physics, Department of Space Science, Beihang University, Beijing
	100191, China\\\vspace{4mm}
	$^2$Department of Physics, Beijing Normal University, Beijing 100875, China\\\vspace{4mm}

	$^3$Key Laboratory of Multiscale Spin Physics (Beijing Normal University), Ministry of Education, Beijing 100875, China\\\vspace{4mm}
	$^4$Peng Huanwu Collaborative Center for Research and Education, Beihang University, Beijing
	100191, China\\\vspace{4mm}
}
\end{center}

\vspace{8mm}

\begin{abstract}

Assuming that the compact source at the Galactic center, Sagittarius A*, is a massive boson star, we fit the near-infrared flare astrometry data. We consider 12 discrete boson star configurations and model the flare as a hotspot on a circular equatorial orbit. The analysis is performed in a Bayesian framework using nested sampling, yielding the marginal posterior distributions of all parameters as well as the Bayesian evidence for each model. For comparison, the same procedure is applied to a Schwarzschild black hole. The resulting Bayesian evidence values differ only marginally between the boson star and black hole cases, and the well-determined mass of Sgr~A* (${\sim}4.296\times 10^6\,M_\odot$) falls within the 68\% highest density interval in every configuration. We conclude that, under current near-infrared astrometric constraints and within the considered parameter ranges, a massive boson star and a Schwarzschild black hole remain statistically indistinguishable as the compact object at the Galactic center.

\end{abstract}

\vfill{
	\footnotesize $\ast$ Corresponding author: minyongguo@bnu.edu.cn
	
	\footnotesize $\dagger$ Corresponding author: hqzhang@buaa.edu.cn

}

\maketitle

\newpage
\baselineskip 18pt

\section{Introduction}\label{sec1}
Sagittarius~A* (Sgr~A*), the nearest supermassive compact object at the Galactic center, lies about 27,000 light-years away with an estimated mass of $\sim4.3 \times 10^{6}M_\odot$, making it an ideal laboratory for strong-field gravity studies. Long-term monitoring of S-star orbits provided precise mass measurements and confirmed relativistic precession \cite{Gillessen:2009ht,Ghez:2008ms,GRAVITY:2018ofz,GRAVITY:2020gka}. Near-infrared interferometry with GRAVITY tracked bright hot spots, revealing a roughly one-hour clockwise orbit and independently verifying the mass \cite{Abuter:2018uum,GRAVITY:2020lpa,GRAVITY:2023avo}. In 2022, the Event Horizon Telescope (EHT) produced the first horizon-scale image, showing a central brightness depression with a surrounding ring, consistent with a magnetically arrested disk model \cite{EventHorizonTelescope:2022wkp,EventHorizonTelescope:2022exc,EventHorizonTelescope:2024hpu}. Extensive theoretical studies have further explored the dynamics, imaging, and relativistic effects of Sgr~A* \cite{EventHorizonTelescope:2023VI,VonFellenberg:2023aa,Johnson:2018aa,Psaltis:2020xus,Zhang:2024lsf,Zhang:2025vyx}.

Although observational and theoretical studies support the black hole interpretation, current data remain indirect reconstructions of gravitational effects or emission-region structures. Current data cannot yet discriminate between a classical black hole and horizonless compact objects, such as bosonic stars \cite{Schunck:2003kk,Jetzer:1991jr,Brito:2015pxa}, gravastars\cite{Mazur:2001fv}, and etc. The latter class of models can equally produce microarcsecond-scale shadows \cite{Kubo:2016ada,Saleem:2024kld,Cardoso:2019rvt,Sengo:2024pwk,Bambi:2025wjx,Vincent:2015xta,Cunha:2018gql,Zhang:2021xhp,Torres:2000dw,Cardoso:2017cqb}, Keplerian motion with periods of about an hour \cite{Fromm:2021flr,Rosa:2025dzq,Broderick:2009ph,Cardoso:2017njb}, and long-period orbital precession \cite{Tamm:2025jrx,Ferreira:2017pth,Kleihaus:2005me}, while naturally circumventing the singularity problem and the information paradox. Exploring horizonless compact objects therefore constitutes not only a necessary cross-check of existing observations, but also a promising avenue toward new physics in the strong-field regime.

Boson stars represent a particularly well-motivated family of horizonless compact objects. They are formed by the gravitational condensation of a large number of integer-spin bosons and possess neither an event horizon nor a spacetime singularity. Their masses and radii are governed by the self-interaction strength of the scalar field and the central field amplitude, and their structure can be solved self-consistently by coupling general relativity with the Klein--Gordon equation. When the self-interaction is sufficiently strong, the compactness of boson stars can rival that of neutron stars or even approach that of black holes, with the effective radius descending close to the Schwarzschild radius $r \approx 2.81\,M$ \cite{Cardoso:2021ehg}, where 
M
M is the boson star mass. The dynamical evolution of boson stars has been investigated in detail, confirming their viability as stable configurations \cite{Liebling:2012fv,DiGiovanni:2020ror}. Moreover, research into boson stars as black hole mimickers has blossomed, encompassing gravitational wave signatures\cite{Yunes:2016jcc,Cardoso:2016oxy,Sennett:2017etc,CalderonBustillo:2020fyi}, accretion dynamics\cite{Torres:2002td,Guzman:2005bs,Guzman:2009zz,Macedo:2013jja,Palenzuela:2017kcg,Olivares:2018abq,Olivares-Sanchez:2024dfh}, and electromagnetic imaging\cite{Herdeiro:2021lwl,He:2025qmq,Rosa:2022tfv,Rosa:2022toh,Rosa:2023qcv,Zhang:2025xnl}. Collectively, these studies demonstrate that, as a representative class of horizonless compact objects, boson stars can effectively replicate the observational signatures of classical black holes in the strong-field regime, underscoring their status as independent astrophysical entities deserving of thorough investigation.

In this study, we assumed Sgr~A$^*$ to be a massive boson star and fitted the near-infrared flare astrometry data within a Bayesian framework. Previously, numerous studies have imposed constraints on the parameters of black holes and hot spots \cite{Xie:2025skg, Aimar:2025uia,Yfantis:2024eab,Antonopoulou:2024qco,Yfantis:2023wsp,Ball:2020jup,Matsumoto:2020wul}. We adopt a method similar to that used in our previous work \cite{Wang:2026teu} to analyze solitonic boson star model. For the two free parameters of the boson star---the coupling parameter $\Lambda$ and the central field amplitude $\psi_0$---we discretely selected 12 distinct configurations. We modeled the flare emission as a hot spot on a circular equatorial orbit and obtained the model centroid positions corresponding to the observational data via general relativistic ray-tracing (GRRT) imaging for each configuration. The posterior distributions were sampled using nested sampling, with a Gaussian likelihood and uniform priors. As a comparison, we applied the identical procedure to the Schwarzschild black hole model. The results show that the Bayesian evidence for the Schwarzschild black hole (SBH) and that for each boson star configuration are of the same order, showing no significant difference; the highest density intervals (HDI) and the equal-tailed $2\sigma$ credible intervals (CI) of the boson star parameters all overlap with the well-established value ranges. This indicates that, within the parameter space considered in our study and under the constraints of current near-infrared astrometry data, Sgr~A$^*$ cannot be statistically distinguished from a black hole to a boson star.

The rest of the paper is organized as follows. In Sec~.\ref{sec2}, we review the numerical solution procedure for boson stars and present the specific configurations adopted in this study. Sec~.\ref{sec3} briefly introduces the radiation model and computational methods used for imaging. Sec~.\ref{sec4} details the setup of the fitting strategy, including the choice of fitting parameters and the construction of the likelihood function. Sec~.\ref{sec5} presents the posterior distributions obtained from sampling and provides a comparative analysis of the imaging characteristics of boson stars and the Schwarzschild black hole. Finally, Sec~.\ref{sec6} summarizes the research and discusses the relevant conclusions together with future prospects.

\section{Boson star solutions}\label{sec2}
In this section, we briefly introduce the massive boson star solutions. The action of the Klein–Gordon–Einstein system is
\begin{equation}
\mathcal{S} = \int d^4x \sqrt{-g} \left[ \frac{R}{2\kappa} - \frac{1}{2} \nabla_a \Phi^* \nabla^a\Phi - V(|\Phi|^2) \right],
\end{equation}
and the potential massive boson stars is
\begin{equation}
V = \mu^2 |\Phi|^2  + \Lambda|\Phi|^4
\end{equation}
where $\Lambda$ is the coupling constant. Observations from GRAVITY indicate that the flare originates outside the innermost stable circular orbit (ISCO) \cite{GRAVITY:2020gka}. At such radii, the spin of the star has a negligible effect on equatorial orbits. Hence, in this work, we adopt a spherically symmetric metric ansatz,
\be
\label{dsansatz}
ds^2 = -A dt^2 + B^{-1} dr^2 + r^2 (d\theta^2+\sin^2\theta d\phi^2),\, 
\ee
and scalar field ansatz is
\be
\label{phiansatz}
\Phi = \psi(r) e^{-i\omega t}.
\ee
Varying the action and substituting the metric and field ansatz yields the equations of motion for the entire system. The explicit forms of these equations and the boundary conditions have been presented in our previous work \cite{Wang:2026teu} and are therefore not repeated here.
\begin{figure}[!htbp]
	\centering
	\includegraphics[width=0.45\textwidth]{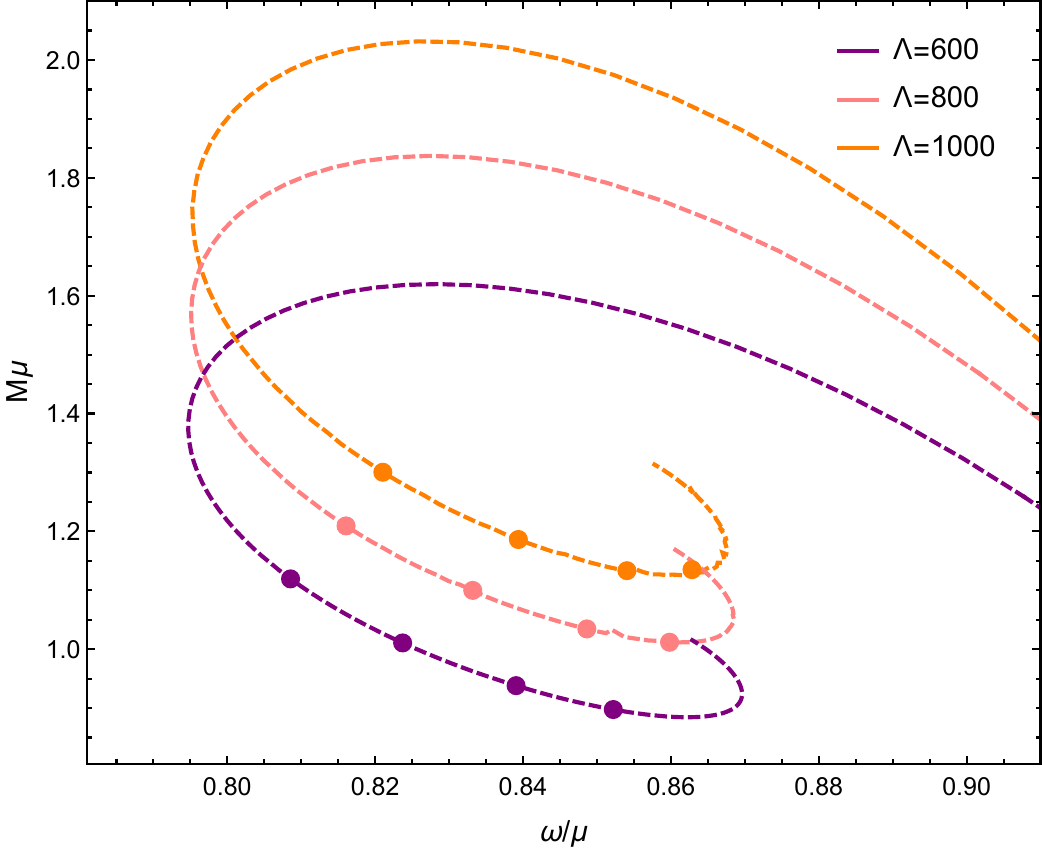}
	\caption{The $M$--$\omega$ curves for different coupling parameters $\Lambda$, with the corresponding $\Lambda$ values indicated in the legend. The boson star configurations adopted in this study are marked by scattered points on the plot.}
	\label{cu}
\end{figure}
The system of differential equations involves four unknown parameters: the boson mass $\mu$, the boundary value $A_0$ of the geometric quantity $A(r)$ at the origin, the field amplitude at the origin $\psi_0$, and the coupling constant $\Lambda$. However, $\mu$ can be absorbed via a global transformation, since the entire system is invariant under the following transformation:
\begin{equation}
	r\rightarrow\mu r,\,A(r)\rightarrow\mu^2A(r),\,V\rightarrow\mu^{-2}V.
\end{equation}
Since we are considering stationary solutions, the field equations do not depend explicitly on the time coordinate $t$. Consequently, the influence of the metric function $A_0$ can be eliminated through a redefinition of $t$. In summary, the solution procedure involves only two free parameters, $\psi_0$ and $\Lambda$, and different boson star configurations are obtained by varying these two parameters. Throughout this work, we restrict ourselves to ground state solutions. When the system is not stiff, the most direct approach is the shooting method: with the boundary conditions imposed, we tune the eigenfrequency $\omega_c$ until the numerical solution exhibits the correct asymptotic behavior at infinity. For each configuration, we focus on two characteristic quantities: the total mass $M$ and the effective radius $R$. The mass $M$ is extracted from the asymptotic fall-off of the metric at infinity, 
\begin{equation}
	M\equiv\lim_{r\rightarrow\infty}\frac{r}{2}(1-B(r)),
\end{equation} 
 the effective radius $R$ is defined as the radial position that encloses 98\% of the total mass of the boson star
 and is defined via
 \begin{equation}
 	B(r)=1-\frac{2m(r)}{r},\quad m(R) = 0.98M.
 \end{equation}

\begin{table}[!htbp]
	\centering
	\renewcommand{\arraystretch}{1.3}
	\begin{tabular}{l|c|c|c|c|c|c}
		\hline\hline
		Model & $\Lambda$ &$\psi_0$ & $\mu M$ & $\mu R$ & $\omega/\mu$ & $R/M$ \\
	    \hline 
		BS1 & 600 & 0.042 & 1.407 & 8.19 & 0.795 & 5.82 \\ 
		\hline 
		BS2 & 600 & 0.052 & 1.235 & 7.340 & 0.798 & 5.94 \\ 
		\hline
		BS3 & 600 & 0.062 & 1.094 & 6.876 & 0.812 & 6.28  \\ 
		\hline
		BS4 & 600 & 0.083 & 0.994 & 6.804 & 0.303 & 6.85 \\ 
		\hline 
		BS5 & 800 & 0.039 & 1.548 & 8.760 & 0.795 & 5.65 \\ \hline 
		BS6 & 800 & 0.049 & 1.338 & 7.842 & 0.803 & 5.86 \\\hline 
		BS7 & 800 & 0.059 & 1.183 & 7.512 & 0.819 & 6.34 \\\hline 
		BS8 & 800 & 0.069 & 1.081 & 7.581 & 0.836 & 7.01 \\\hline 
		BS9 & 1000 & 0.036 & 1.690 & 9.390 & 0.795 & 5.55 \\\hline 
		BS10 & 1000 & 0.046 & 1.443 & 8.402 & 0.807 & 5.82 \\\hline 
		BS11 & 1000 & 0.056 & 1.273 & 8.173 & 0.825 & 6.42 \\\hline 
		BS12 & 1000 & 0.066 & 1.171 & 8.333 & 0.842 & 7.11 \\
	
		\hline\hline
	\end{tabular}
	\caption{Parameters for different boson star model.}
	\label{para}
\end{table}

In this study, we fit boson star configurations with different central field amplitude $\psi_0$ and coupling parameters $\Lambda$. We select three different coupling parameters, $\Lambda = 600, 800, 1000$, corresponding to different self-interaction strengths in the scalar field. Previous studies have shown that the constraint results are closely related to the compactness of the star. Therefore, for each given self-interaction strength, we use a grid-based sampling approach to find the central field amplitude value $\psi_0^{\rm min}$ that makes the star most compact (i.e., with the smallest effective radius), and then take $\psi_0^{\rm min} + 0.1, +0.2, +0.3$ successively to examine the influence of the $\psi_0$ on the constraint results for a fixed coupling constant. All adopted configurations are listed in Table~\ref{para}. Figure~\ref{cu} shows the $M$--$\omega$ curves for $\Lambda = 600, 800, 1000$, with the selected configurations marked as scatter points. It should be noted that this paper focuses only on the trend of the constraint results under different parameters, and therefore the stability of the boson stars is not discussed.

\section{Emission model and imaging process}\label{sec3}
In this section, we briefly outline the procedure for obtaining the model-predicted observable: the centroid position. For imaging, we employ the ray-tracing method\cite{Chen:2024jkm}. The observer’s screen is discretized into a grid of pixels, and the geodesics are integrated backward in time. Each pixel corresponds to a null geodesic, and the observed intensity $I_o$ for each geodesic is determined by the radiative transfer equation
\bea
I_o = \nu_o \int g^2 \, J \, d\lambda,
\eea
where $g \equiv \nu_o/\nu_e$, is the redshift factor relating the observed frequency $\mu_o$ to the emitted frequency
$\mu_e$, $\lambda$ is the affine parameter of the geodesic, and $J$ is the emission coefficient, which depends on the light source.  In our work, we consider the hot spot as a Gaussian source, and the emission coefficient is
\bea
J\propto e^{-\frac{|\vec{\boldsymbol{r}}_p-\vec{\boldsymbol{r}}_\text{hs}|^2}{2s^2}},
\eea
where $s$ is the Gaussian radius of the hot spot, in our work we set $s=0.3M$, and $|\vec{\boldsymbol{r}}_p-\vec{\boldsymbol{r}}_\text{hs}|$ represents the spatial distance between the emission point and the center of the hot spot, where $\vec{\boldsymbol{r}}_p$ is the spatial coordinate of the emission point and $\vec{\boldsymbol{r}}_\text{hs}$ is the position of the hot spot. We consider a circular-orbit hot spot, whose four-velocity and angular velocity are given by the following formulae
\begin{equation}
	p^\mu=\zeta(1,0,0,\Omega),\,\,\,\, \Omega_{\pm} =\pm\sqrt{\frac{-\partial_r A(r)}{2r}},
\end{equation}
where $\zeta$ is the normalization factor, $\Omega$ is the orbital angular velocity, $A(r)$ is a geometric quantity given by the solution procedure introduced in the previous section, and the ``$\pm$" represent prograde and retrograde orbits.

Since GRAVITY measures the brightness centroid, we image the hotspot at a given time to produce a snapshot and then compute its centroid, which serves as the model prediction corresponding to the observation. To do so, we first define the flux of a single pixel.
\begin{equation}
F(i,j) = I_o(i,j) S_0 \cos \Psi_{\text{in}}(i,j),
\end{equation}
where $I_o(i,j)$ is the intensity of the pixel $(i,j)$, $S_0$ represents the pixel size and 
$\Psi_{\text{in}}$ denotes the incident angle of the light ray relative to the imaging plane. 
For the camera model we used, the incident angle $\Psi_{\text{in}}$ takes the form \cite{Huang:2024wpj}
\begin{equation}
\Psi_{\text{in}}(i,j) = 2 \arctan \left(\frac{2}{N} \tan \left( \frac{\alpha_{\text{fov}}}{2} \right) \sqrt{ \left( i - \frac{N+1}{2} \right)^2 + \left( j - \frac{N+1}{2} \right)^2 } \right),
\end{equation}
where $N$ is the total number of pixels in each column or row, and $ \alpha_{\text{fov}} $ is the field of view angle. Then, the centroid of emission $\left(X_c,\,Y_c\right)$ can be obtained as the weighted average of the position with the flux serving as the weight, namely,
\begin{equation}
X_c = \frac{\sum\limits_{i,j}X(i,j) F(i,j)}{\sum\limits_{i,j} F(i,j)},\,\,\,\,\,Y_c = \frac{\sum\limits_{i,j}Y(i,j) F(i,j)}{\sum\limits_{i,j} F(i,j)},
\end{equation}

\section{Fitting process}\label{sec4}
In this section, we describe the parameter estimation procedure in detail. We use the combined astrometric data released by the GRAVITY collaboration, which are obtained by averaging the astrometric measurements of four individual flare events. According to the GRAVITY analysis, the four flares share the same set of orbital parameters; averaging the data therefore helps suppress the impact of individual-event uncertainties on the inferred global parameters. In this work, we consider the following parameters:

\begin{equation}
    \boldsymbol{\Theta}=(r_\text{hs},\, \theta_o,\,\text{PA},\, \phi_0\,,M),
\end{equation}
 in which, $r_\text{hs}$ and $\phi_0$ denote the orbital radius and initial azimuth of the hot spot, respectively; these two parameters vary between individual flares. The inclination angle $\theta_o$, position angle PA, and mass $M$ of Sgr~A* are global parameters shared by all events. We restrict our analysis to these five parameters. Additional factors, such as the $x$--$y$ coordinate offsets and the finite size of the hot spot, have a negligible impact on the results and are therefore omitted, consistent with the approach adopted by GRAVITY.
 \begin{table}[h]
 	\centering
 	\begin{tabular}{c|c|c}
 		\hline\hline
 		Parameter & Meaning & Range  \\
 		\hline
 		$r_{\text{hs}}$ & orbital radius  & $[4M, 12M]$  \\
 		$\theta_o$ & inclination & $[90^\circ,\,180^\circ]$  \\
 		$\phi_0$ & initial azimuth & $[0, \,360^\circ]$  \\
 		PA & position angle & $[0, \,180^\circ]$ \\
 		$M$ & Sgr A* mass & $[1\times10^6 M_\odot, \,11\times10^6 M_\odot]$  \\
 		\hline
 	\end{tabular}
 	\caption{The prior range for each parameter}
 	\label{range}
 \end{table}
 Under the Bayesian framework, the distribution function of the parameters can be obtained as the product of the likelihood function and the prior probability. The likelihood function $\mathcal{L}(\boldsymbol{\Theta})$  was defined in a standard way as
 \begin{equation}
 	\log\mathcal{L}(\boldsymbol{\Theta}) \equiv-\frac{1}{2}\sum_i \left[ \frac{\left( X_c^i(\boldsymbol{\Theta}) - X_o^i \right)^2 }{ \sigma_{X^i}^2 }+\frac{\left( Y_c^i(\boldsymbol{\Theta}) - Y_o^i \right)^2 }{ \sigma_{Y^i}^2 } \right] ,
 \end{equation}
 the subscript ``c'' denotes quantities computed from the model parameters, while the subscript ``o'' refers to the observed quantities (i.e., the combined astrometric data). The superscript $i$ labels the astrometric epoch $t_i$. The observational uncertainties in the $x$ and $y$ directions are denoted by $\sigma_{X^i}$ and $\sigma_{Y^i}$, respectively. We note that the log-likelihood function $\log\mathcal{L}(\boldsymbol{\Theta})$ is constructed as a residual-based measure, so that a larger log-likelihood value corresponds to a smaller residual. Accordingly, after obtaining the sampling results, we adopt the maximum of $\mathcal{L}$ as the global optimal parameter.

We continue to introduce the other part of the parameter distribution function, namely the prior function. In this paper, we consider a uniform function, the prior function is 
\[
\pi(\boldsymbol{\Theta}) = 
\begin{cases}
	1, & a \leq \boldsymbol{\Theta} \leq b, \\
	0, & \text{\text{otherwise}}.
\end{cases}
\]
Here, $a$ and $b$ denote the range of the prior parameter, this prior function means that the probability density is uniform within the interval $[a,b]$ and zero outside this interval. The range of each prior parameter is given in Tab.~\ref{range}.

By integrating the distribution function, the resulting normalization constant is the Bayesian evidence $\mathcal{Z}$,
\begin{equation}
\mathcal{Z}=\int_{\Omega_{\boldsymbol{\Theta}}}\mathcal{L}(\boldsymbol{\Theta})\pi(\boldsymbol{\Theta})d\boldsymbol{\Theta}.
\end{equation}
The Bayesian evidence is defined by the integral over the full parameter space $\Omega_{\boldsymbol{\Theta}}$ and essentially quantifies the marginal likelihood of a model given the data, accounting for the structure of the parameter space and the prior. It thus serves as a quantitative measure of the overall plausibility of the model. We use the Python package \texttt{Dynesty}\cite{Speagle2020_dynesty} to perform nested sampling within the Bayesian framework. \texttt{Dynesty} provides both the posterior distributions of the parameters and the Bayesian evidence for different models.
 \begin{table}[h]
 	\centering
 	\begin{tabular}{c|c|c}
 		\hline\hline
 		Parameter & Grid & Total number  \\
 		\hline
 		orbital radius $r_{\text{hs}}$ & 4, 4.5, 5... 11.5, 12 & $17$  \\
 		inclination $\theta_o$ & $90^\circ,\,95^\circ,100^\circ...175^\circ,\,180^\circ$ & $19$  \\
 		\hline
 	\end{tabular}
 	\caption{The grid of $r_\text{hs}$ and $\theta_o$.}
 	\label{grid}
 \end{table}

 In practice, we construct the likelihood function through the following steps. For a given boson star configuration, each point in the $(r_\text{hs},\, \theta_o)$ subspace determines a hotspot movie, from which a preliminary centroid trajectory is obtained. The parameters $\phi_0$, PA, and $M$ then correspond to a time shift, a rotation, and a scaling of this trajectory, respectively. Together, these five parameters define the model prediction. Notably, computing the hotspot movie requires only $r_\text{hs}$ and $\theta_o$. We therefore discretize these two parameters over a grid (see Tab.~\ref{grid}) and precompute the corresponding centroid trajectories. Subsequently, we interpolate over $(r_\text{hs},\, \theta_o, t)$ and incorporate the remaining parameters ($\phi_0$, PA, $M$) to evaluate the full likelihood function. Since high-dimensional interpolation is required, we use the \texttt{RegularGridInterpolator} from the Python \texttt{SciPy} package\cite{Virtanen2020_scipy}, which performs well in multidimensional settings.

\section{Results}\label{sec5}

We used the combined astrometry dataset to constrain the parameters of massive boson stars. For comparison, we also constrained the parameters of the Schwarzschild black hole over the same parameter range. In the process of computing the hotspot movie, the image resolution was set to $64^2$. At this resolution, the captured imaging details differ little from those at high resolution. Furthermore, during the sampling process, we set the number of active points \texttt{nlive} $=9000$ and set the sampling termination condition to $d\log \mathcal{Z} = 1$.

\begin{figure}[!htbp]
	\centering
	\includegraphics[width=0.30\textwidth]{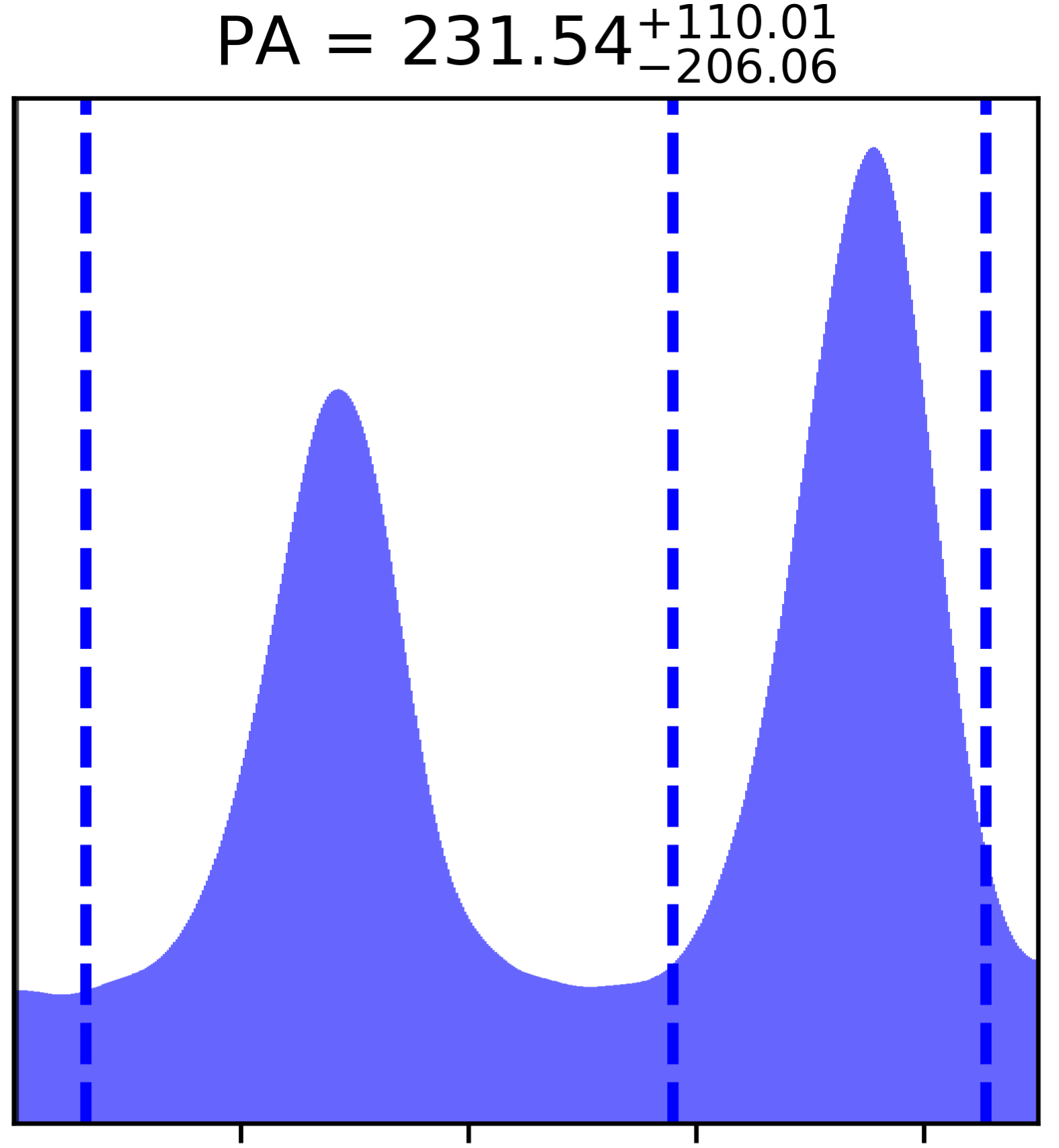}
	\caption{Posterior distribution of the position angle for BS5, where the dashed lines mark the CI and the median.}
	\label{PA}
\end{figure}

\subsection{Sample result}
Using nested sampling, we inferred the imaging parameters for the 12 boson star models. From the posterior samples, we identified the maximum-likelihood point and computed the corresponding $\chi^2$ statistic; the parameters at that point are taken as the best-fit values. The results are summarized in Tab.~\ref{dataall}. For each configuration, the first row lists the best-fit parameters, and the second row reports the $95\%$ ($2\sigma$) equal-tail CI and the medians, all calculated with the \texttt{dyplot} function provided by \texttt{Dynesty}.

\begin{figure}[!htbp]
	\centering
	\includegraphics[width=0.99\textwidth]{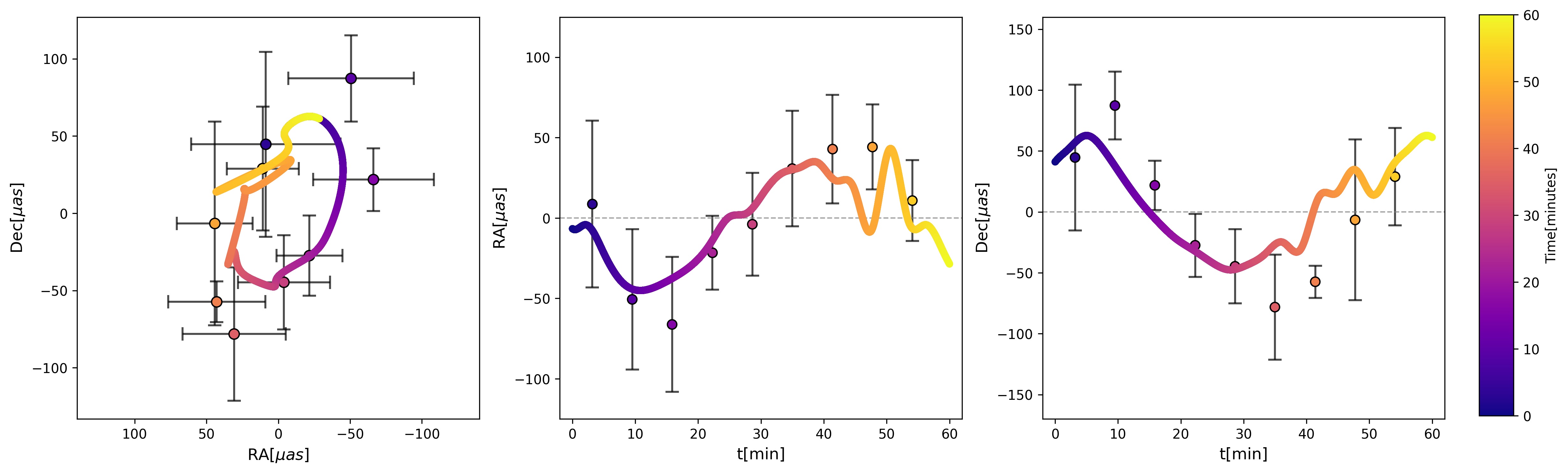}
	\caption{The best fit of BS1 model for combined data. Left: the centroid motion. Middle and Right: the Right Ascension and Declination on the sky as a function of time.}
	\label{chires}
\end{figure}
We first examine the CI of the mass parameter. For the SBH model, the mass is constrained to $5.09^{+4.90}_{-2.34}\times 10^6\,M_\odot$. This CI not only includes the S-star orbital measurement of $(4.297 \pm 0.0012) \times 10^6\,M_\odot$, but also agrees well with the results from the EHT ($4.0^{+1.1}_{-0.6}\times 10^6\,M_\odot$) and GRAVITY ($4.17^{+1.11}_{-0.83}\times 10^6\,M_\odot$). However, due to the specific choice of priors and the markedly right-skewed posterior, the median mass of $5.09\times 10^6\,M_\odot$ is noticeably offset relative to these reference values. Compared with the SBH case, the boson star models tend to yield larger median masses, a consequence of their even more strongly skewed posteriors. Nevertheless, the overall CIs of the two classes of models remain highly overlapping. For $r_{\text{hs}}$, all constraints fall within similar intervals. When the results are considered collectively, a clear trend emerges: smaller inferred masses correspond to larger $r_{\text{hs}}$ values. The parameters $\theta_{\text{o}}$ and $\phi_0$ are also constrained to similar intervals across models. In contrast, the position angle PA exhibits larger fluctuations, which stem from the bimodal structure of its posterior distribution. Fig.~\ref{PA} shows the posterior of PA for configuration BS1, where the bimodality is clearly visible. We attribute such scatter in the angular parameters primarily to the absence of polarimetric constraints; the QU loop would normally provide much tighter bounds on these quantities.

Turning to the best-fit results, we note that the $\theta_{\text{o}}$ values are all located near the peak of their posterior. We also observe that some best-fit points drift toward extremely large masses, a situation that is typically accompanied by very small $r_{\text{hs}}$; we will analyze this phenomenon in detail in a later section. The best-fit PA jumps between two angular ranges, corresponding exactly to the two peaks of its posterior. The best-fit $\phi_0$ shows no obvious systematic trend, as its value is determined jointly by multiple parameters. Fig.~\ref{chires} displays the model with the smallest $\chi^2$ of BS1. The three panels compare the overall orbital trajectory, as well as the temporal evolution of RA and Dec, with the observational data.

Finally, we analyze the log-Bayesian evidence $\log \mathcal{Z}$. None of the boson star configurations we considered show a significant difference compared to the Schwarzschild black hole model. Even for BS1, the configuration with the largest discrepancy, the Bayesian factor $\Delta\log \mathcal{Z}$ relative to SBH is only $0.16$, which is far below $1$. According to the Jeffreys scale, $\Delta\log \mathcal{Z} < 1$ constitutes ``weak'' evidence, meaning the models cannot be meaningfully distinguished.This indicates that, within the parameter ranges considered in this work and given the current data, all boson star models are statistically equivalent to the SBH, and the data do not favor any particular model. Furthermore, within the boson star family itself, a mild trend is discernible over our chosen parameter range: the results tend to prefer larger $\Lambda$ and larger $\psi_0$.

\begin{table}[!htbp]
    \centering
    \renewcommand{\arraystretch}{1.3}
    \begin{tabular}{|l|l|c|c|c|c|c|c|}
        \hline\hline
        \multicolumn{2}{|c|}{Model} & $r_\text{hs}[M]$ & $\theta_o[^\circ]$ & $\phi_0[^\circ]$ & PA$[^\circ]$ & $M[10^6M_\odot]$ & $\log\mathcal{Z}$ \\
        \hline
        \multirow{2}{*}{BS1} 
        & Fit & 7.39 & 125.40 & 55.91 & 311.08 & 5.30 & \multirow{2}{*}{$-16.260\pm0.078$} \\ \cline{2-7}
        & HDI & $7.55^{+4.17}_{-3.01}$ & $134.50^{+42.98}_{-39.30}$ & $174.62^{+176.62}_{-165.67}$ & $134.50^{+42.98}_{-39.30}$ & $5.67^{+4.74}_{-2.84}$&\\\hline
        \multirow{2}{*}{BS2} 
        & Fit & 7.04 & 117.95 & 111.09 & 301.64
      	& 5.36 & \multirow{2}{*}{$-16.395\pm0.076$} \\\cline{2-7} 
        & HDI & $7.97^{+3.75}_{-3.37}$ & $130.76^{+46.76}_{-36.28}$ & $178.75^{+172.14}_{-169.72}$ & $257.77^{+80.11}_{-229.39}$ & $5.12^{+4.96}_{-2.32}$ & \\
        \hline
        \multirow{2}{*}{BS3} 
        & Fit & 4.44 & 114.77 & 271.99 & 316.24 & 10.85 & \multirow{2}{*}{$-16.346\pm0.071$} \\ \cline{2-7}
        & HDI & $7.80^{+3.94}_{-3.37}$ & $135.33^{+42.23}_{-38.98}$ & $172.04^{+179.61}_{-163.46}$ & $203.82^{+139.15}_{-181.07}$ & $5.46^{+5.08}_{-2.67}$ & \\ 
        \hline
        \multirow{2}{*}{BS4} 
        & Fit & 6.12 & 119.88 & 99.47 & 115.19 & 7.69 & \multirow{2}{*}{$-16.423\pm0.075$} \\ \cline{2-7}
        & HDI & $7.88^{+3.85}_{-3.40}$ & $137.04^{+40.44}_{-41.20}$& $176.03^{+174.55}_{-166.97}$ & $214.00^{+130.31}_{-193.48}$  & $5.29^{+5.18}_{-2.48}$ & \\ 
        \hline
        \multirow{2}{*}{BS5} 
        & Fit & 4.49 & 117.98 & 287.72 & 306.45 & 9.48 & \multirow{2}{*}{$-16.290\pm0.077$} \\ \cline{2-7}
        & HDI & $7.56^{+4.16}_{-3.06}$ & $134.35^{+43.33}_{-39.15}$ & $179.59^{+170.78}_{-170.40}$ & $231.54^{+110.01}_{-260.06}$ & $5.71^{+4.79}_{-2.88}$ & \\ 
        \hline
         \multirow{2}{*}{BS6} 
         & Fit & 5.55 & 109.81 & 36.10 & 314.40 & 9.93 & \multirow{2}{*}{$-16.360\pm0.076$} \\ \cline{2-7}
        & HDI & $7.75^{+3.99}_{-3.23}$ & $137.37^{+40.36}_{-41.89}$ & $177.50^{+173.17}_{-167.59}$ & $212.02^{+132.16}_{-193.80}$ & $5.61^{+4.89}_{-2.82}$ & \\  
        \hline
         \multirow{2}{*}{BS7} 
        & Fit & 5.04 & 112.20 & 156.99 & 297.03 & 10.56 & \multirow{2}{*}{$-16.360\pm0.074$} \\ \cline{2-7}
	      & HDI & $7.79^{+3.95}_{-3.30}$ & $135.13^{+42.29}_{-39.15}$ & $176.90^{+173.35}_{-167.81}$ & $207.89^{+136.22}_{-187.35}$ & $5.43^{+5.10}_{-2.65}$ & \\
        \hline
         \multirow{2}{*}{BS8} 
        & Fit & 7.96 & 106.03 & 255.65 & 115.90 & 6.45 & \multirow{2}{*}{$-16.421\pm0.074$} \\ \cline{2-7}
        & HDI & $7.82^{+3.91}_{-3.42}$ & $137.56^{+39.87}_{-41.13}$& $176.29^{+174.47}_{-167.19}$ & $203.82^{+141.38}_{-182.62}$  & $5.40^{+5.16}_{-2.62}$ & \\ 
        \hline
         \multirow{2}{*}{BS9} 
        & Fit & 4.49 & 120.80 & 292.86 & 301.23 & 9.60 & \multirow{2}{*}{$-16.295\pm0.073$}  \\ \cline{2-7}
        & HDI & $7.61^{+4.09}_{-3.11}$ & $134.41^{+43.43}_{-38.94}$ & $180.79^{+168.97}_{-171.01}$ & $228.73^{+113.30}_{-203.23}$ & $5.69^{+4.80}_{-2.89}$ & \\ 
        \hline
         \multirow{2}{*}{BS10} 
        & Fit & 5.04 & 106.50 & 164.69 & 295.64 & 10.52 & \multirow{2}{*}{$-16.382\pm0.077$} \\ \cline{2-7}
        & HDI & $7.78^{+3.95}_{-3.26}$ & $136.75^{+40.81}_{-41.04}$ & $177.53^{+172.35}_{-167.13}$ & $214.02^{+129.41}_{-193.38}$ & $5.56^{+4.96}_{-2.78}$ & \\ 
        \hline
         \multirow{2}{*}{BS11} 
        & Fit & 6.63 & 114.23 & 188.81 & 307.63 & 7.54 & \multirow{2}{*}{$-16.366\pm0.077$} \\ \cline{2-7}
        & HDI & $7.75^{+3.99}_{-3.27}$ & $136.33^{+41.39}_{-39.96}$ & $176.81^{+174.40}_{-167.88}$ & $201.34^{+141.36}_{-179.66}$ & $5.48^{+5.07}_{-2.70}$ & \\ 
        \hline
         \multirow{2}{*}{BS12} 
        & Fit & 8.27 & 112.16 & 224.90 & 115.98 & 6.24 & \multirow{2}{*}{$-16.421\pm0.076$} \\ \cline{2-7}
        & HDI & $7.85^{+3.89}_{-3.39}$ & $137.67^{+39.93}_{-41.53}$ & $179.27^{+171.67}_{-170.11}$ & $221.35^{+123.57}_{-199.79}$ & $5.33^{+5.12}_{-2.54}$ &  \\ 
        \hline
        \multirow{2}{*}{SBH} 
        & Fit & 10.36 & 113.37 & 87.40 & 261.822 & 4.10 & \multirow{2}{*}{$-16.398\pm0.075$} \\ \cline{2-7}
        & HDI & $8.03^{+3.71}_{-3.23}$ & $130.51^{+46.58}_{-35.42}$ & $176.95^{+174.52}_{-168.49}$ & $257.32^{+85.32}_{-230.14}$ & $5.09^{+4.90}_{-2.34}$ &  \\ 
        \hline\hline
    \end{tabular}
    \caption{Comparison between best-fit parameter values and posterior median values (95\% HDI) of combined data for different compact object models.}
    \label{dataall}
\end{table}

\subsection{Posterior Distribution of $M$}

In the previous section, we found that the Bayesian evidence for the SBH and that for the boson star are comparable, indicating that---based on the flare astrometry alone---the boson star model explains the data as well as the black hole model. We also noticed that the boson star model tends to push the inferred Galactic center mass toward larger values. Motivated by this, we present a dedicated analysis of the mass parameter in this section. Note that all mass intervals reported earlier were obtained using the \texttt{dyplot} tool from \texttt{Dynesty}, which provides equal-tailed CI. Since most of the posteriors are skewed, CI do not adequately capture their shapes. We therefore recompute the mass intervals using highest density intervals (HDI), specifically via the \texttt{hdi} function of the \texttt{arviz} library \cite{Kumar2022_arviz}.
\begin{figure}[!htbp]
	\centering
	\includegraphics[width=0.9\textwidth]{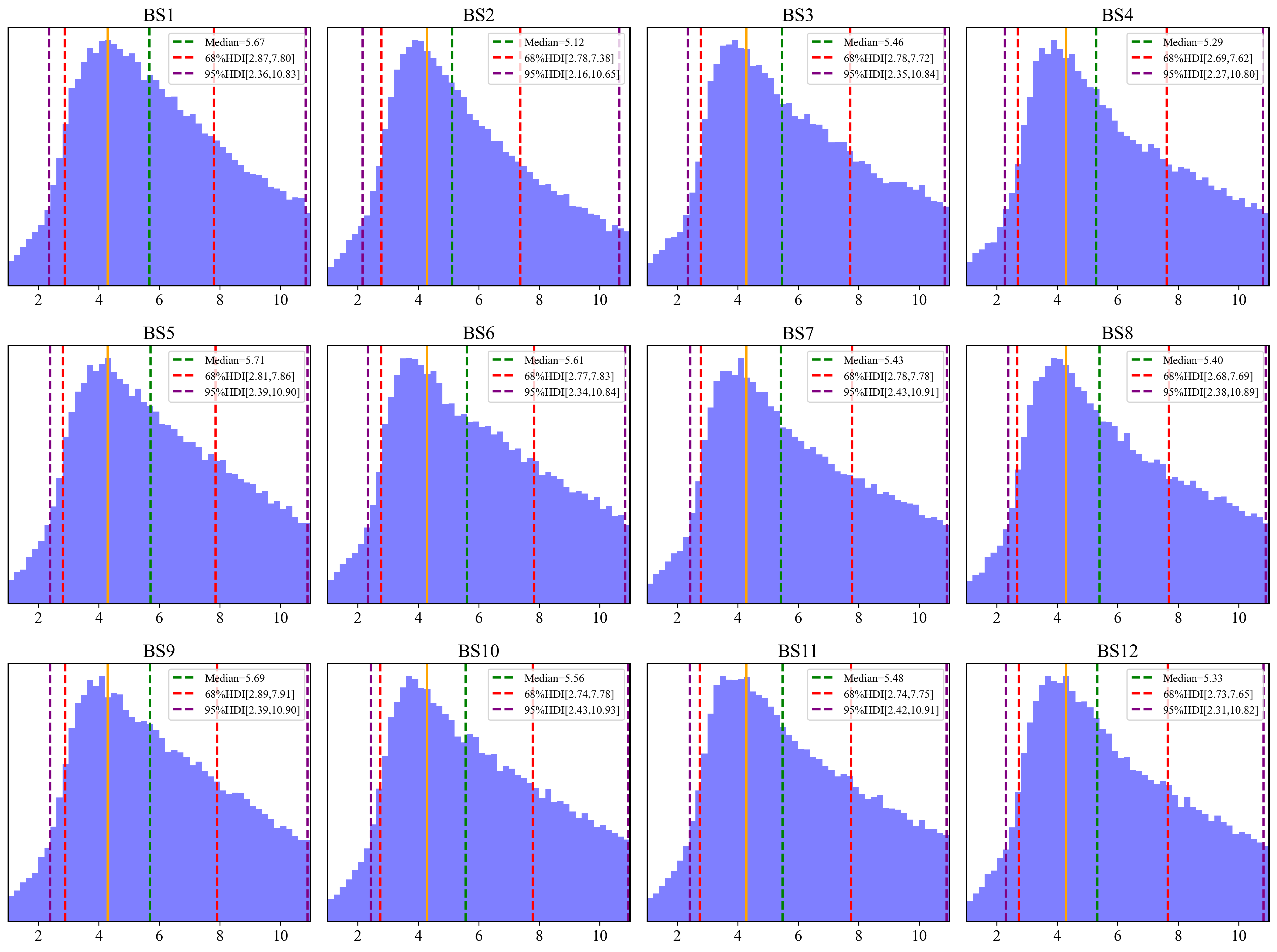}
	\caption{Posterior distribution plots of the mass parameter for all boson star configurations. In each panel, the green line marks the median, the red line marks the 68\% HDI, and the purple line marks the 95\% HDI.}
	\label{hdi}
\end{figure}
\begin{figure}[!htbp]
	\centering
	\includegraphics[width=0.4\textwidth]{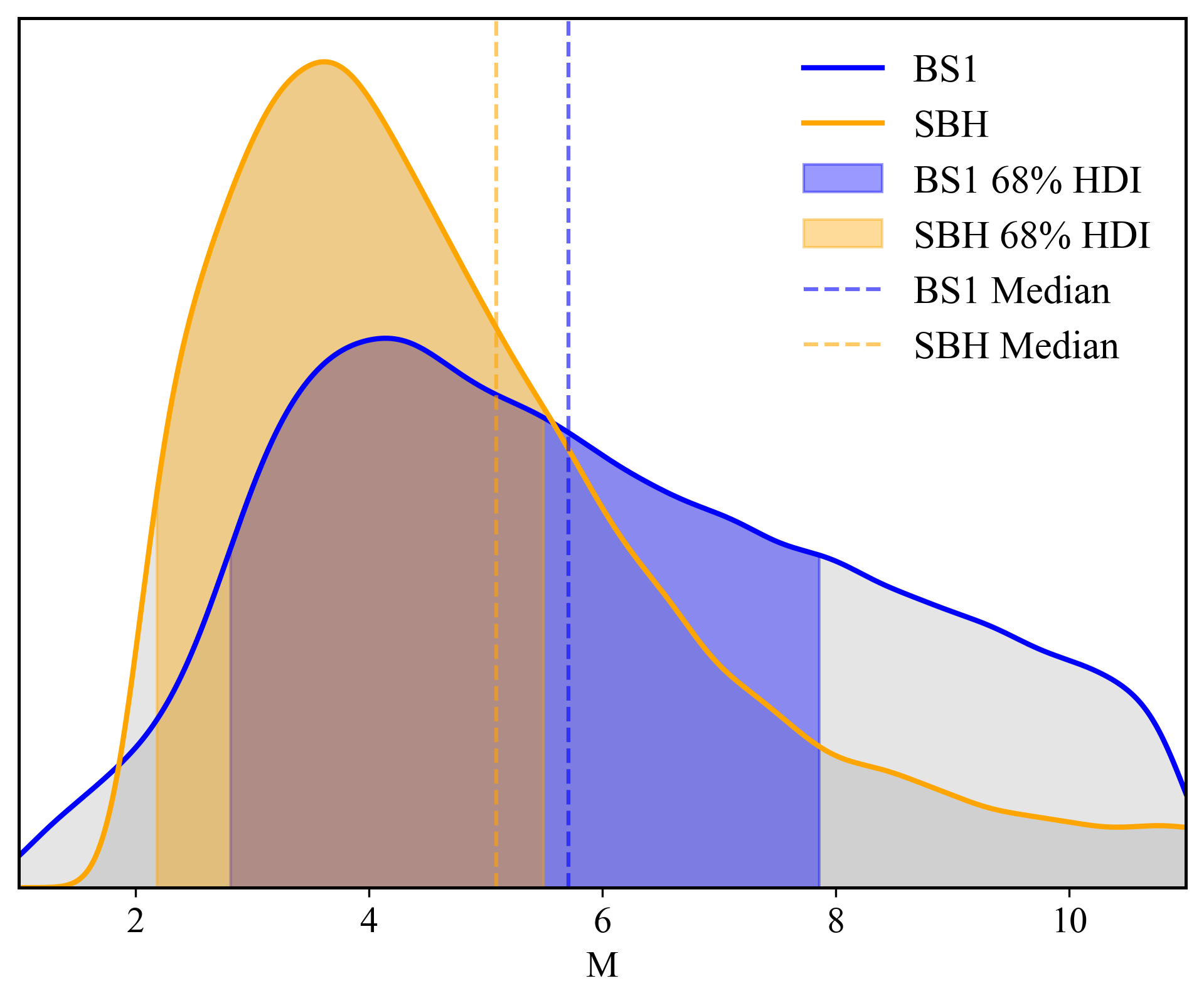}
	\caption{The marginal posterior distributions of the mass for BS1 and SBH. The yellow curve corresponds to SBH, and the blue curve corresponds to BS1. The areas below each curve filled with the corresponding colors represent the 68\% HDI, while the parts outside the HDI are filled in gray. The two dashed lines mark the medians of the respective distributions.}
	\label{sbc}
\end{figure}

Figure~\ref{hdi} displays the posterior distributions of the mass parameter for all boson star configurations. In each panel, the green, red, and purple dashed lines mark the median, 68\% HDI, and 95\% HDI, respectively. The horizontal axis represents the inferred mass in units of \(10^6\,M_\odot\), and the vertical axis shows the probability density. The yellow solid line indicates the S-star mass measurement of Sgr~A*, \((4.297 \pm 0.012) \times 10^6\,M_\odot\). For every configuration, the posterior peak lies close to this reference value, demonstrating that the mass constraints obtained from the boson star models are fully compatible with the independent mass determination of the Galactic center compact object. Although slight differences exist among the posteriors of different configurations, all models reproduce the Sgr~A* mass reasonably well, which supports the possibility that boson stars could be viable candidate central objects.
Furthermore, each posterior retains a relatively high probability density at the high-mass end, whereas the low-mass tail decays rapidly. This asymmetry leads to a noticeable shift of the median away from the posterior peak.

We now examine the differences between the boson star model and the SBH. Since the mass posteriors of all boson star configurations share a similar shape, we select BS1 as a representative case for a direct comparison with SBH. Fig.~\ref{sbc} shows the marginal posterior distributions of BS1 (blue) and SBH (yellow); the 68\% HDI regions are shaded in the corresponding colors, and the medians are indicated by dashed lines. The SBH posterior is also skewed, but less than that of BS1, and is more sharply concentrated around the peak. Consequently, the 68\% HDI of SBH is only about half as wide as that of BS1, and the probability density near the peak is appreciably higher. In other words, compared with SBH, BS1 yields a broader mass interval and a flatter posterior, reflecting greater uncertainty and weaker constraining power on the mass parameter.

\begin{figure}[!htbp]
	\centering
	\includegraphics[width=0.7\textwidth]{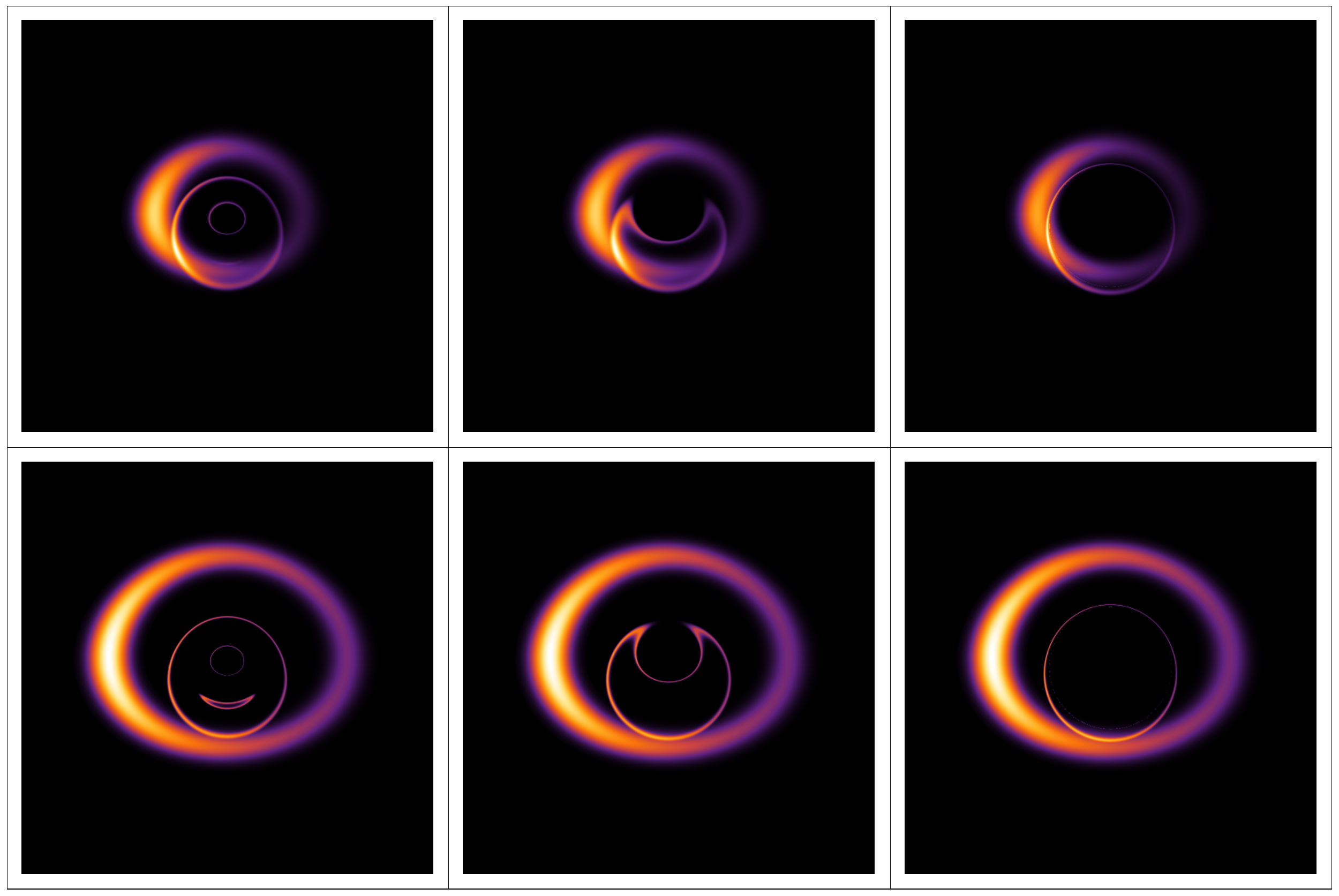}
	\caption{Time-averaged images of BS1, BS4, and the SBH at an inclination angle $\theta_o = 135^\circ$. The columns, from left to right, correspond to BS1, BS4, and SBH, respectively. The top and bottom rows display results for hotspot orbital radii $r_{\text{hs}} = 5M$ and $9M$.}
	\label{img}
\end{figure}

\subsection{From the image perspective}
To investigate the previously noted tendency of boson star models to yield larger and more skewed mass estimates, we analyze this behavior in conjunction with the imaging results in this section. Figure~\ref{img} displays the time-averaged images of BS1, BS4, and the SBH for hot spot orbital radii $r_{\text{hs}} = 5M$ and $9M$, at an inclination angle $\theta_o = 135^\circ$. The columns, from left to right, correspond to BS1, BS4, and SBH, respectively.

\begin{figure}[!htbp]
	\centering
	\includegraphics[width=0.76\textwidth]{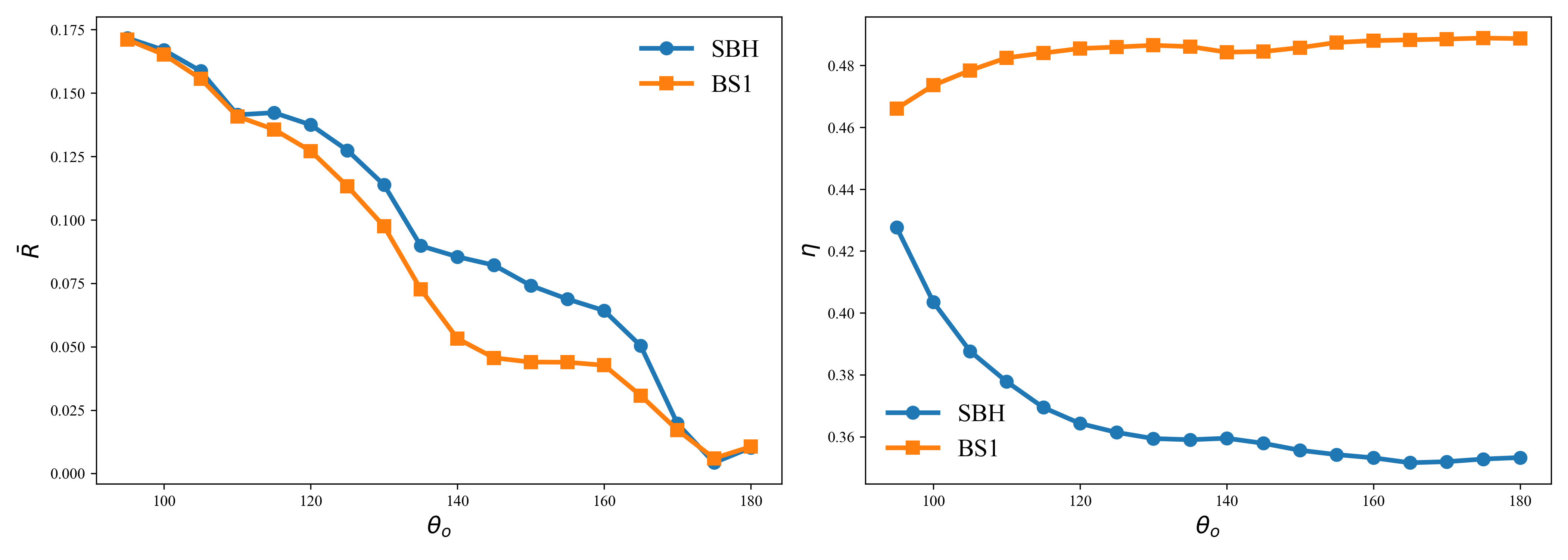}
	\caption{Left panel: $\bar{R}$ as a function of $\theta_o$. Right panel: $\eta$ as a function of $\theta_o$. The yellow curves represent BS1, and the blue curves represent the SBH. The adopted grid points are indicated by markers.}
	\label{ri}
\end{figure}
Several features are evident from the images. For the SBH, the time-averaged image is dominated by the primary hotspot image and the photon ring. In contrast, because boson stars possess no event horizon, photons can traverse the stellar interior and produce so-called plunge-through images. When computing the centroid trajectory, these plunge-through image systematically shift the brightness centroid inward. During the fitting process, however, the model must reproduce the orbital radius required by the astrometric data. To compensate for the inward-shifted centroid, the fit tends to infer a larger mass, since a larger mass globally enlarges the image scale. A larger mass also implies a longer orbital period at a given orbital radius; therefore, the data favor smaller hotspot radii with faster angular velocities. This interplay explains why the mass posterior of the boson star is systematically shifted toward higher values compared with the SBH. Moreover, the occasional appearance of best‑fit solutions with extremely large masses and very small radii is a direct consequence of the same mechanism.

In the time-averaged images, the high‑order images—defined as those formed by photons that cross the equatorial plane at least twice—shrink the centroid trajectory; the plunge‑through images mentioned above constitute a subset of these. In what follows, we present a quantitative analysis of the high‑order images.
The centroid $(\bar{X}_c,\bar{Y}_c)$ of the high-order component is computed as
\begin{equation}
	\bar{X}_c = \frac{\sum_{i,j} X(i,j)\, \bar{F}(i,j)}{\sum_{i,j} \bar{F}(i,j)},\qquad
	\bar{Y}_c = \frac{\sum_{i,j} Y(i,j)\, \bar{F}(i,j)}{\sum_{i,j} \bar{F}(i,j)},
\end{equation}
where $\bar{F}(i,j)$ denotes the flux of the high-order image at pixel $(i,j)$. From the centroid we define the mean radius $\bar{R} = \sqrt{\bar{X}_c^2 + \bar{Y}_c^2}$ of the high-order component in each time-averaged image. The left panel of Fig.~\ref{ri} displays $\bar{R}$ as a function of $\theta_o$; the yellow curve corresponds to BS1, the blue curve to the Schwarzschild black hole, and the markers indicate the selected angles. For both models, $\bar{R}$ decreases with increasing $\theta_o$, and the mean radius of BS1 is consistently smaller than that of the Schwarzschild black hole. We further define the flux fraction $\eta$ of the high-order image relative to the total image as
\begin{equation}
	\eta = \frac{\sum_{i,j} \bar{F}(i,j)}{\sum_{i,j} F(i,j)},
\end{equation}
and its angular variation is shown in the right panel of Fig.~\ref{ri}, following the same color coding. As the $\theta_o$ increases, $\eta$ gradually decreases for the Schwarzschild black hole, whereas it slowly increases for BS1. Earlier analysis indicates that the high-order image affects the inferred mass $M$ by shrinking the overall centroid track. Comparing the two panels, we find that for both BS1 and SBH the mean radius $\bar{R}$ decreases with $\theta_o$, with a more pronounced decline for BS1; a smaller $\bar{R}$ implies a stronger inward displacement of the centroid. At the same time, the flux fraction $\eta$ for BS1 slowly grows, further amplifying the track contraction driven by the reduction of $\bar{R}$. In other words, the contraction induced by the high-order component varies with $\theta_o$, ultimately producing a broad, right-skewed posterior distribution of $M$ for BS1. For SBH, on the other hand, the flux fraction decreases with viewing angle, and the two effects partially cancel, yielding a more compact and concentrated mass posterior.

\section{Summary and discussion}\label{sec6}

In this study, we model Sgr~A* as a massive boson star and fit the near-infrared astrometric flare data released by the GRAVITY collaboration, assuming a circular, equatorial hotspot orbit. From the boson star parameter space we discretely sample 12 configurations defined by different combinations of the coupling constant $\Lambda$ and the central field amplitude $\psi_0$. For each configuration we simulate a hotspot movie at a grid of orbital radii $r_{\text{hs}}$ and inclination angles $\theta_o$, and extract the centroid trajectory as the model prediction to be compared with the astrometry. The full set of model parameters---$r_{\text{hs}}$, $\theta_o$, position angle PA, initial azimuth $\phi_0$, and the mass of Sgr~A*, $M$---are constrained via nested sampling; the centroid curves are rotated, time‑shifted and scaled to incorporate PA, $\phi_0$ and $M$, respectively. For comparison we apply exactly the same fitting procedure to a SBH.

The nested sampling yields the log-Bayesian evidence $\log\mathcal{Z}$ for each model. No significant evidence difference is found between any BS configuration and the SBH. We further present the $2\sigma$ equal‑tail CIs for all parameters, and for the mass $M$ we additionally compute the $68\%$ and $95\%$ HDIs. Both the HDIs and the CIs of all models comfortably contain the well‑established Galactic center mass $(4.297\pm0.012)\times10^{6}\,M_\odot$. All marginal mass posteriors are noticeably skewed, with an offset between the median and the peak, yet the peaks consistently lie near the literature value.  
Based on these results we conclude that, under the constraints of the GRAVITY near‑infrared astrometry alone and for the range of boson star parameters considered here, one cannot statistically distinguish whether Sgr~A* is a black hole or a massive boson star---both remain equally viable candidates for the central compact object.

Comparing the posteriors in more detail, we find that the BS mass distributions are systematically shifted toward higher values and exhibit more pronounced skewness than the SBH case; their HDIs and CIs are also broader. We explain this behavior by inspecting the image structure.  
Unlike the SBH, a boson star possesses no event horizon, so light can traverse its interior and produce ``plunge‑through'' photons that have no counterpart in black hole images. These extra photons appear near the image center, pulling the brightness centroid inward and reducing the apparent orbital displacement. As we have shown previously, a smaller centroid displacement cannot easily match the astrometrically required orbit scale; the fit therefore compensates by inflating $M$, which globally stretches the image. A larger mass, however, lengthens the orbital period and forces the solution toward a smaller hotspot radius $r_{\text{hs}}$.  
Moreover, plunge‑through photons contribute a significant fraction of the total flux, and the balance between the direct and indirect images of the BS varies appreciably with $\theta_o$. This introduces a degeneracy between $M$ and $\theta_o$. Altogether, these effects produce a mass posterior that is flatter, more skewed, and broader than that of the SBH.

The present study explores only a small slice of the massive boson star parameter space, focusing on a limited set of discrete $(\Lambda,\psi_0)$ pairs. In the future we plan to upgrade our GRRT imaging framework to scan the full parameter space systematically, enabling a more comprehensive assessment of the compatibility between boson star models and observations. Extending the analysis to rotating boson stars is another natural direction that may bring new insight into mass constraints and imaging signatures.  
On the observational side, the angular parameters remain only weakly constrained by the astrometry alone; incorporating polarimetric data is essential. Polarization observables are highly sensitive to the geometry and can efficiently break the degeneracies that persist in astrometry‑only fits. We leave these improvements for future work and anticipate that more accurate forward modeling and higher‑resolution imaging will significantly advance the investigation of boson stars as candidates for the Galactic center compact object.

\section*{Acknowledgments}
We are grateful to Zhenyu Zhang for discussions. This work was partially supported by the National Natural Science Fundation of China (Grant No. 12575048 and 12175008). M. Guo is also supported by the BNU Tang Scholar.

\bibliographystyle{utphys}
\bibliography{gravityflare}

@article{Gillessen:2009ht,
    author = "Gillessen, S. and Eisenhauer, F. and Fritz, T. K. and Bartko, H. and Dodds-Eden, K. and Pfuhl, O. and Ott, T. and Genzel, R.",
    title = "{The orbit of the star S2 around SgrA* from VLT and Keck data}",
    eprint = "0910.3069",
    archivePrefix = "arXiv",
    primaryClass = "astro-ph.GA",
    doi = "10.1088/0004-637X/707/2/L114",
    journal = "Astrophys. J. Lett.",
    volume = "707",
    pages = "L114--L117",
    year = "2009"
}

@article{Ghez:2008ms,
		author = "Ghez, A. M. and others",
		title = "{Measuring Distance and Properties of the Milky Way's Central Supermassive Black Hole with Stellar Orbits}",
		eprint = "0808.2870",
		archivePrefix = "arXiv",
		primaryClass = "astro-ph",
		doi = "10.1086/592738",
		journal = "Astrophys. J.",
		volume = "689",
		pages = "1044--1062",
		year = "2008"
}

@article{GRAVITY:2018ofz,
    author = "Abuter, R. and others",
    collaboration = "GRAVITY",
    title = "{Detection of the gravitational redshift in the orbit of the star S2 near the Galactic centre massive black hole}",
    eprint = "1807.09409",
    archivePrefix = "arXiv",
    primaryClass = "astro-ph.GA",
    doi = "10.1051/0004-6361/201833718",
    journal = "Astron. Astrophys.",
    volume = "615",
    pages = "L15",
    year = "2018"
}

@article{Kubo:2016ada,
		author = "Kubo, Tomohiro and Sakai, Nobuyuki",
		title = "{Gravitational lensing by gravastars}",
		doi = "10.1103/PhysRevD.93.084051",
		journal = "Phys. Rev. D",
		volume = "93",
		number = "8",
		pages = "084051",
		year = "2016"
}

@inbook{Olivares-Sanchez:2024dfh,
    author = "Olivares-S{\'a}nchez, H{\'e}ctor R. and Kocherlakota, Prashant and Herdeiro, Carlos A. R.",
    title = "{GRMHD Simulations of~Accretion Onto Exotic Compact Objects}",
    eprint = "2408.09893",
    archivePrefix = "arXiv",
    primaryClass = "astro-ph.HE",
    doi = "10.1007/978-981-97-8522-3_15",
    year = "2025"
}

@article{Psaltis:2020xus,
  author = {Psaltis, Dimitrios and Wex, Norbert and Kramer, Michael},
  title = {A Quantitative Test of the No-Hair Theorem with Sgr A* Using Stellar Orbits and Pulsar Timing},
  journal = {The Astrophysical Journal},
  volume = {889},
  pages = {52},
  year = {2020},
  doi = {10.3847/1538-4357/ab5b3d}
}

@article{Johnson:2018aa,
  author = {Johnson, Michael D. and Fish, Vincent L. and Doeleman, Sheperd S. and et al.},
  title = {The size, shape, and scattering of Sagittarius A* at 86 GHz: First VLBI with ALMA},
  journal = {The Astrophysical Journal},
  volume = {865},
  pages = {104},
  year = {2018},
  doi = {10.3847/1538-4357/aad5c4}
}

@article{VonFellenberg:2023aa,
  author = {Von Fellenberg, S. and Lu, R. S. and Psaltis, D. and et al.},
  title = {Testing General Relativity with Sgr A*: Relativistic Photon Ring Predictions from GRMHD Models},
  journal = {Monthly Notices of the Royal Astronomical Society},
  volume = {518},
  pages = {1234--1250},
  year = {2023},
  doi = {10.1093/mnras/stac2874}
}

@article{EventHorizonTelescope:2023VI,
  author = {{Event Horizon Telescope Collaboration}},
  title = {The Physics of the Accretion Flow and Shadow of Sgr A*: Theoretical Interpretation of EHT Observations},
  journal = {The Astrophysical Journal Letters},
  volume = {950},
  pages = {L12},
  year = {2023},
  doi = {10.3847/2041-8213/acf46a}
}

@article{Chen:2024jkm,
    author = "Chen, Bin and Hou, Yehui and Song, Yu and Zhang, Zhenyu",
    title = "{Polarization patterns of the hot spots plunging into a Kerr black hole}",
    eprint = "2407.14897",
    archivePrefix = "arXiv",
    primaryClass = "astro-ph.HE",
    doi = "10.1103/PhysRevD.111.083045",
    journal = "Phys. Rev. D",
    volume = "111",
    number = "8",
    pages = "083045",
    year = "2025"
}

@article{Zhang:2024lsf,
    author = "Zhang, Zhenyu and Hou, Yehui and Guo, Minyong and Chen, Bin",
    title = "{Imaging thick accretion disks and jets surrounding black holes}",
    eprint = "2401.14794",
    archivePrefix = "arXiv",
    primaryClass = "astro-ph.HE",
    doi = "10.1088/1475-7516/2024/05/032",
    journal = "JCAP",
    volume = "05",
    pages = "032",
    year = "2024"
}

@article{Zhang:2025vyx,
    author = "Zhang, Zhenyu and Hou, Yehui and Guo, Minyong and Mizuno, Yosuke and Chen, Bin",
    title = "{Autocorrelation signatures in time-resolved black hole flare images: Secondary peaks and convergence structure}",
    eprint = "2503.17200",
    archivePrefix = "arXiv",
    primaryClass = "astro-ph.HE",
    doi = "10.1103/zmnz-p2rs",
    journal = "Phys. Rev. D",
    volume = "112",
    number = "8",
    pages = "083024",
    year = "2025"
}

@article{Mazur:2001fv,
    author = "Mazur, Pawel O. and Mottola, Emil",
    title = "{Gravitational Condensate Stars: An Alternative to Black Holes}",
    eprint = "gr-qc/0109035",
    archivePrefix = "arXiv",
    reportNumber = "LA-UR-01-5067, LA-UR-01-5067",
    doi = "10.3390/universe9020088",
    journal = "Universe",
    volume = "9",
    number = "2",
    pages = "88",
    year = "2023"
}

@article{Kumar2022_arviz,
  title={ArviZ: A unified library for exploratory analysis of Bayesian models in Python},
  author={Kumar, Rahul and Carroll, Chris and Hartikainen, Aleksi and Martin, Oliver and Corander, Jukka and J{\"o}nsson, Petter and Lindsten, Fredrik and Paananen, Tommi and Simola, Jouni and Virtanen, Sami and others},
  journal={Journal of Open Source Software},
  volume={7},
  number={76},
  pages={2889},
  year={2022},
  doi={10.21105/joss.02889}
}

@article{Virtanen2020_scipy,
  title={SciPy 1.0: Fundamental Algorithms for Scientific Computing in Python},
  author={Virtanen, Pauli and Gommers, Ralf and Oliphant, Travis E. and Haberland, Matt and Reddy, Tyler and Cournapeau, David and Burovski, Evgeni and Peterson, Pearu and Weckesser, Warren and Bright, Jonathan and others},
  journal={Nature Methods},
  volume={17},
  pages={261--272},
  year={2020},
  doi={10.1038/s41592-019-0686-2}
}

@article{Speagle2020_dynesty,
  title={Dynesty: a dynamic nested sampling package for estimating Bayesian posteriors and evidences},
  author={Speagle, Joshua S.},
  journal={Monthly Notices of the Royal Astronomical Society},
  volume={493},
  number={3},
  pages={3132--3158},
  year={2020},
  publisher={Oxford University Press},
  doi={10.1093/mnras/staa278}
}

@article{He:2025qmq,
    author = "He, Ke-Jian and Li, Guo-Ping and Yang, Chen-Yu and Zeng, Xiao-Xiong",
    title = "{The observation image of a soliton boson star illuminated by various accretions}",
    eprint = "2502.16623",
    archivePrefix = "arXiv",
    primaryClass = "gr-qc",
    doi = "10.1088/1475-7516/2025/10/003",
    journal = "JCAP",
    volume = "10",
    pages = "003",
    year = "2025"
}

@article{Abuter:2018uum,
    author = "Abuter, R. and others",
    title = "{Detection of orbital motions near the last stable circular orbit of the massive black hole SgrA*}",
    eprint = "1810.12641",
    archivePrefix = "arXiv",
    primaryClass = "astro-ph.GA",
    doi = "10.1051/0004-6361/201834294",
    journal = "Astron. Astrophys.",
    volume = "618",
    year = "2018"
}

@article{Sengo:2024pwk,
    author = "Sengo, Ivo and Cunha, Pedro V. P. and Herdeiro, Carlos A. R. and Radu, Eugen",
    title = "{The imitation game reloaded: effective shadows of dynamically robust spinning Proca stars}",
    eprint = "2402.14919",
    archivePrefix = "arXiv",
    primaryClass = "gr-qc",
    doi = "10.1088/1475-7516/2024/05/054",
    journal = "JCAP",
    volume = "05",
    pages = "054",
    year = "2024"
}

@article{Brito:2015pxa,
    author = "Brito, Richard and Cardoso, Vitor and Herdeiro, Carlos A. R. and Radu, Eugen",
    title = "{Proca stars: Gravitating Bose{\textendash}Einstein condensates of massive spin 1 particles}",
    eprint = "1508.05395",
    archivePrefix = "arXiv",
    primaryClass = "gr-qc",
    doi = "10.1016/j.physletb.2015.11.051",
    journal = "Phys. Lett. B",
    volume = "752",
    pages = "291--295",
    year = "2016"
}

@article{Jetzer:1991jr,
    author = "Jetzer, Philippe",
    title = "{Boson stars}",
    reportNumber = "ZU-TH-25-91",
    doi = "10.1016/0370-1573(92)90123-H",
    journal = "Phys. Rept.",
    volume = "220",
    pages = "163--227",
    year = "1992"
}

@article{Schunck:2003kk,
    author = "Schunck, Franz E. and Mielke, Eckehard W.",
    title = "{General relativistic boson stars}",
    eprint = "0801.0307",
    archivePrefix = "arXiv",
    primaryClass = "astro-ph",
    doi = "10.1088/0264-9381/20/20/201",
    journal = "Class. Quant. Grav.",
    volume = "20",
    pages = "R301--R356",
    year = "2003"
}

@article{Yfantis:2023wsp,
    author = "Yfantis, A. I. and Mo{\'s}cibrodzka, M. A. and Wielgus, M. and Vos, J. T. and Jimenez-Rosales, A.",
    title = "{Fitting the light curves of Sagittarius A* with a hot-spot model - Bayesian modeling of QU loops in the millimeter band}",
    eprint = "2310.07762",
    archivePrefix = "arXiv",
    primaryClass = "astro-ph.HE",
    doi = "10.1051/0004-6361/202348230",
    journal = "Astron. Astrophys.",
    volume = "685",
    pages = "A142",
    year = "2024"
}

@article{Vincent:2015xta,
    author = "Vincent, F. H. and Meliani, Z. and Grandclement, P. and Gourgoulhon, E. and Straub, O.",
    title = "{Imaging a boson star at the Galactic center}",
    eprint = "1510.04170",
    archivePrefix = "arXiv",
    primaryClass = "gr-qc",
    doi = "10.1088/0264-9381/33/10/105015",
    journal = "Class. Quant. Grav.",
    volume = "33",
    number = "10",
    pages = "105015",
    year = "2016"
}

@article{Kleihaus:2005me,
    author = "Kleihaus, Burkhard and Kunz, Jutta and List, Meike",
    title = "{Rotating boson stars and Q-balls}",
    eprint = "gr-qc/0505143",
    archivePrefix = "arXiv",
    doi = "10.1103/PhysRevD.72.064002",
    journal = "Phys. Rev. D",
    volume = "72",
    pages = "064002",
    year = "2005"
}

@article{Ferreira:2017pth,
    author = "Ferreira, Miguel C. and Macedo, Caio F. B. and Cardoso, Vitor",
    title = "{Orbital fingerprints of ultralight scalar fields around black holes}",
    eprint = "1710.00830",
    archivePrefix = "arXiv",
    primaryClass = "gr-qc",
    doi = "10.1103/PhysRevD.96.083017",
    journal = "Phys. Rev. D",
    volume = "96",
    number = "8",
    pages = "083017",
    year = "2017"
}

@article{Wang:2026teu,
    author = "Wang, Xiangyu and Zhang, Zhenyu and Zhang, Hai-Qing and Guo, Minyong and Chen, Bin",
    title = "{Testing solitonic boson star interpretations of Sagittarius A* with near-infrared flare astrometry}",
    eprint = "2604.21883",
    archivePrefix = "arXiv",
    primaryClass = "astro-ph.HE",
    month = "4",
    year = "2026"
}

@article{Rosa:2025dzq,
    author = "Rosa, Jo{\~a}o Lu{\'\i}s and Aimar, Nicolas and Tamm, Hanna Liis",
    title = "{Polarimetry imprints of exotic compact objects: Solitonic boson stars}",
    eprint = "2504.02472",
    archivePrefix = "arXiv",
    primaryClass = "gr-qc",
    doi = "10.1103/h8ln-psjr",
    journal = "Phys. Rev. D",
    volume = "111",
    number = "12",
    pages = "124036",
    year = "2025"
}

@article{Aimar:2025uia,
    author = "Aimar, Nicolas and Rosa, Jo{\~a}o Lu{\'\i}s and Tamm, Hanna Liis and Garcia, Paulo",
    title = "{Sagittarius A* near-infrared flares polarization as a probe of space-time I: Non-rotating exotic compact objects}",
    eprint = "2506.23931",
    archivePrefix = "arXiv",
    primaryClass = "astro-ph.HE",
    doi = "10.1051/0004-6361/202557669",
    journal = "Astron. Astrophys.",
    volume = "708",
    pages = "A379",
    year = "2026"
}

@article{Tamm:2025jrx,
    author = "Tamm, Hanna Liis and Aimar, Nicolas and Rosa, Jo{\~a}o Lu{\'\i}s",
    title = "{Polarimetry imprints of exotic compact objects: relativistic fluid spheres and gravastars}",
    eprint = "2509.20344",
    archivePrefix = "arXiv",
    primaryClass = "gr-qc",
    month = "9",
    year = "2025"
}

@article{Rosa:2023qcv,
    author = "Rosa, Jo{\~a}o Lu{\'\i}s and Macedo, Caio F. B. and Rubiera-Garcia, Diego",
    title = "{Imaging compact boson stars with hot spots and thin accretion disks}",
    eprint = "2303.17296",
    archivePrefix = "arXiv",
    primaryClass = "gr-qc",
    doi = "10.1103/PhysRevD.108.044021",
    journal = "Phys. Rev. D",
    volume = "108",
    number = "4",
    pages = "044021",
    year = "2023"
}

@inproceedings{Bambi:2025wjx,
    author = "Bambi, Cosimo and others",
    title = "{Black hole mimickers: from theory to observation}",
    eprint = "2505.09014",
    archivePrefix = "arXiv",
    primaryClass = "gr-qc",
    month = "5",
    year = "2025"
}

@article{Saleem:2024kld,
    author = "Saleem, Rabia and Aslam, M. Israr and Shahid, Shokaib",
    title = "{Observational signatures of charged rotating traversable wormhole: shadows and light rings with different accretions}",
    doi = "10.1140/epjc/s10052-024-12853-z",
    journal = "Eur. Phys. J. C",
    volume = "84",
    number = "5",
    pages = "480",
    year = "2024"
}

@article{Cardoso:2019rvt,
    author = "Cardoso, Vitor and Pani, Paolo",
    title = "{Testing the nature of dark compact objects: a status report}",
    eprint = "1904.05363",
    archivePrefix = "arXiv",
    primaryClass = "gr-qc",
    doi = "10.1007/s41114-019-0020-4",
    journal = "Living Rev. Rel.",
    volume = "22",
    number = "1",
    pages = "4",
    year = "2019"
}

@article{GRAVITY:2020gka,
    author = "Abuter, R. and others",
    collaboration = "GRAVITY",
    title = "{Detection of the Schwarzschild precession in the orbit of the star S2 near the Galactic centre massive black hole}",
    eprint = "2004.07187",
    archivePrefix = "arXiv",
    primaryClass = "astro-ph.GA",
    doi = "10.1051/0004-6361/202037813",
    journal = "Astron. Astrophys.",
    volume = "636",
    pages = "L5",
    year = "2020"
}

@article{EventHorizonTelescope:2024hpu,
    author = "Akiyama, Kazunori and others",
    collaboration = "Event Horizon Telescope",
    title = "{First Sagittarius A* Event Horizon Telescope Results. VII. Polarization of the Ring}",
    doi = "10.3847/2041-8213/ad2df0",
    journal = "Astrophys. J. Lett.",
    volume = "964",
    number = "2",
    pages = "L25",
    year = "2024"
}

@article{EventHorizonTelescope:2022exc,
    author = "Akiyama, Kazunori and others",
    collaboration = "Event Horizon Telescope",
    title = "{First Sagittarius A* Event Horizon Telescope Results. IV. Variability, Morphology, and Black Hole Mass}",
    eprint = "2311.08697",
    archivePrefix = "arXiv",
    primaryClass = "astro-ph.HE",
    reportNumber = "FERMILAB-PUB-22-423-PPD",
    doi = "10.3847/2041-8213/ac6736",
    journal = "Astrophys. J. Lett.",
    volume = "930",
    number = "2",
    pages = "L15",
    year = "2022"
}

@article{EventHorizonTelescope:2022wkp,
    author = "Akiyama, Kazunori and others",
    collaboration = "Event Horizon Telescope",
    title = "{First Sagittarius A* Event Horizon Telescope Results. I. The Shadow of the Supermassive Black Hole in the Center of the Milky Way}",
    eprint = "2311.08680",
    archivePrefix = "arXiv",
    primaryClass = "astro-ph.HE",
    doi = "10.3847/2041-8213/ac6674",
    journal = "Astrophys. J. Lett.",
    volume = "930",
    number = "2",
    pages = "L12",
    year = "2022"
}

@article{GRAVITY:2023avo,
    author = "Abuter, R. and others",
    collaboration = "GRAVITY",
    title = "{Polarimetry and astrometry of NIR flares as event horizon scale, dynamical probes for the mass of Sgr A*}",
    eprint = "2307.11821",
    archivePrefix = "arXiv",
    primaryClass = "astro-ph.GA",
    doi = "10.1051/0004-6361/202347416",
    journal = "Astron. Astrophys.",
    volume = "677",
    pages = "L10",
    year = "2023"
}

@article{GRAVITY:2020lpa,
    author = {Baub{\"o}ck, M. and others},
    collaboration = "GRAVITY",
    title = "{Modeling the orbital motion of Sgr A*{\textquoteright}s near-infrared flares}",
    eprint = "2002.08374",
    archivePrefix = "arXiv",
    primaryClass = "astro-ph.HE",
    doi = "10.1051/0004-6361/201937233",
    journal = "Astron. Astrophys.",
    volume = "635",
    pages = "A143",
    year = "2020"
}

@article{DiGiovanni:2020ror,
    author = "Di Giovanni, Fabrizio and Sanchis-Gual, Nicolas and Cerd{\'a}-Dur{\'a}n, Pablo and Zilh{\~a}o, Miguel and Herdeiro, Carlos and Font, Jos{\'e} A. and Radu, Eugen",
    title = "{Dynamical bar-mode instability in spinning bosonic stars}",
    eprint = "2010.05845",
    archivePrefix = "arXiv",
    primaryClass = "gr-qc",
    doi = "10.1103/PhysRevD.102.124009",
    journal = "Phys. Rev. D",
    volume = "102",
    number = "12",
    pages = "124009",
    year = "2020"
}

@article{Liebling:2012fv,
    author = "Liebling, Steven L. and Palenzuela, Carlos",
    title = "{Dynamical boson stars}",
    eprint = "1202.5809",
    archivePrefix = "arXiv",
    primaryClass = "gr-qc",
    doi = "10.1007/s41114-023-00043-4",
    journal = "Living Rev. Rel.",
    volume = "26",
    number = "1",
    pages = "1",
    year = "2023"
}

@article{Cardoso:2021ehg,
    author = "Cardoso, Vitor and Macedo, Caio F. B. and Maeda, Kei-ichi and Okawa, Hirotada",
    title = "{ECO-spotting: looking for extremely compact objects with bosonic fields}",
    eprint = "2112.05750",
    archivePrefix = "arXiv",
    primaryClass = "gr-qc",
    doi = "10.1088/1361-6382/ac41e7",
    journal = "Class. Quant. Grav.",
    volume = "39",
    number = "3",
    pages = "034001",
    year = "2022"
}

@article{Antonopoulou:2024qco,
    author = "Antonopoulou, Eleni and Nathanail, Antonios",
    title = "{Parameter study for hot spot trajectories around SgrA*}",
    eprint = "2405.10115",
    archivePrefix = "arXiv",
    primaryClass = "astro-ph.HE",
    doi = "10.1051/0004-6361/202450571",
    journal = "Astron. Astrophys.",
    volume = "690",
    pages = "A240",
    year = "2024"
}

@article{Yfantis:2024eab,
    author = "Yfantis, A. I. and Wielgus, M. and Mo{\'s}cibrodzka, M. A.",
    title = "{Hot spots around Sagittarius A* - Joint fits to astrometry and polarimetry}",
    eprint = "2408.07120",
    archivePrefix = "arXiv",
    primaryClass = "astro-ph.HE",
    doi = "10.1051/0004-6361/202451884",
    journal = "Astron. Astrophys.",
    volume = "691",
    pages = "A327",
    year = "2024"
}

@article{Ball:2020jup,
    author = {Ball, David and {\"O}zel, Feryal and Christian, Pierre and Chan, Chi-Kwan and Psaltis, Dimitrios},
    title = "{A Plasmoid model for the Sgr A* Flares Observed With Gravity and CHANDRA}",
    eprint = "2005.14251",
    archivePrefix = "arXiv",
    primaryClass = "astro-ph.HE",
    doi = "10.3847/1538-4357/abf8ae",
    journal = "Astrophys. J.",
    volume = "917",
    number = "1",
    pages = "8",
    year = "2021"
}

@article{Matsumoto:2020wul,
    author = "Matsumoto, Tatsuya and Chan, Chi-Ho and Piran, Tsvi",
    title = "{The origin of hotspots around Sgr A*: Orbital or pattern motion?}",
    eprint = "2004.13029",
    archivePrefix = "arXiv",
    primaryClass = "astro-ph.HE",
    doi = "10.1093/mnras/staa2095",
    journal = "Mon. Not. Roy. Astron. Soc.",
    volume = "497",
    number = "2",
    pages = "2385--2392",
    year = "2020"
}

@article{Xie:2025skg,
    author = "Xie, Fengting and Zhu, Qing-Hua and Li, Xin",
    title = "{Investigating non-Keplerian motion in flare events with astrometric data}",
    eprint = "2507.07411",
    archivePrefix = "arXiv",
    primaryClass = "astro-ph.HE",
    month = "7",
    year = "2025"
}

@article{Rosa:2022toh,
    author = "Rosa, Jo{\~a}o Lu{\'\i}s and Garcia, Paulo and Vincent, Fr{\'e}d{\'e}ric H. and Cardoso, Vitor",
    title = "{Observational signatures of hot spots orbiting horizonless objects}",
    eprint = "2205.11541",
    archivePrefix = "arXiv",
    primaryClass = "gr-qc",
    doi = "10.1103/PhysRevD.106.044031",
    journal = "Phys. Rev. D",
    volume = "106",
    number = "4",
    pages = "044031",
    year = "2022"
}

@article{Macedo:2013jja,
    author = "Macedo, Caio F. B. and Pani, Paolo and Cardoso, Vitor and Crispino, Lu{\'\i}s C. B.",
    title = "{Astrophysical signatures of boson stars: quasinormal modes and inspiral resonances}",
    eprint = "1307.4812",
    archivePrefix = "arXiv",
    primaryClass = "gr-qc",
    doi = "10.1103/PhysRevD.88.064046",
    journal = "Phys. Rev. D",
    volume = "88",
    number = "6",
    pages = "064046",
    year = "2013"
}

@article{Palenzuela:2017kcg,
    author = "Palenzuela, Carlos and Pani, Paolo and Bezares, Miguel and Cardoso, Vitor and Lehner, Luis and Liebling, Steven",
    title = "{Gravitational Wave Signatures of Highly Compact Boson Star Binaries}",
    eprint = "1710.09432",
    archivePrefix = "arXiv",
    primaryClass = "gr-qc",
    doi = "10.1103/PhysRevD.96.104058",
    journal = "Phys. Rev. D",
    volume = "96",
    number = "10",
    pages = "104058",
    year = "2017"
}

@article{Sennett:2017etc,
    author = "Sennett, Noah and Hinderer, Tanja and Steinhoff, Jan and Buonanno, Alessandra and Ossokine, Serguei",
    title = "{Distinguishing Boson Stars from Black Holes and Neutron Stars from Tidal Interactions in Inspiraling Binary Systems}",
    eprint = "1704.08651",
    archivePrefix = "arXiv",
    primaryClass = "gr-qc",
    doi = "10.1103/PhysRevD.96.024002",
    journal = "Phys. Rev. D",
    volume = "96",
    number = "2",
    pages = "024002",
    year = "2017"
}

@article{CalderonBustillo:2020fyi,
    author = "Calder{\'o}n Bustillo, Juan and Sanchis-Gual, Nicolas and Torres-Forn{\'e}, Alejandro and Font, Jos{\'e} A. and Vajpeyi, Avi and Smith, Rory and Herdeiro, Carlos and Radu, Eugen and Leong, Samson H. W.",
    title = "{GW190521 as a Merger of Proca Stars: A Potential New Vector Boson of $8.7\times 10^{-13}$  eV}",
    eprint = "2009.05376",
    archivePrefix = "arXiv",
    primaryClass = "gr-qc",
    reportNumber = "LIGO DCC:P-2000353",
    doi = "10.1103/PhysRevLett.126.081101",
    journal = "Phys. Rev. Lett.",
    volume = "126",
    number = "8",
    pages = "081101",
    year = "2021"
}

@article{Cardoso:2016oxy,
    author = "Cardoso, Vitor and Hopper, Seth and Macedo, Caio F. B. and Palenzuela, Carlos and Pani, Paolo",
    title = "{Gravitational-wave signatures of exotic compact objects and of quantum corrections at the horizon scale}",
    eprint = "1608.08637",
    archivePrefix = "arXiv",
    primaryClass = "gr-qc",
    doi = "10.1103/PhysRevD.94.084031",
    journal = "Phys. Rev. D",
    volume = "94",
    number = "8",
    pages = "084031",
    year = "2016"
}

@article{Yunes:2016jcc,
    author = "Yunes, Nicolas and Yagi, Kent and Pretorius, Frans",
    title = "{Theoretical Physics Implications of the Binary Black-Hole Mergers GW150914 and GW151226}",
    eprint = "1603.08955",
    archivePrefix = "arXiv",
    primaryClass = "gr-qc",
    doi = "10.1103/PhysRevD.94.084002",
    journal = "Phys. Rev. D",
    volume = "94",
    number = "8",
    pages = "084002",
    year = "2016"
}

@article{Fromm:2021flr,
    author = "Fromm, C. M. and Mizuno, Y. and Younsi, Z. and Olivares, H. and Porth, O. and De Laurentis, M. and Falcke, H. and Kramer, M. and Rezzolla, L.",
    title = "{Using space-VLBI to probe gravity around Sgr A*}",
    eprint = "2101.08618",
    archivePrefix = "arXiv",
    primaryClass = "astro-ph.HE",
    doi = "10.1051/0004-6361/201937335",
    journal = "Astron. Astrophys.",
    volume = "649",
    pages = "A116",
    year = "2021"
}

@article{Guzman:2009zz,
    author = "Guzman, F. S. and Rueda-Becerril, J. M.",
    title = "{Spherical boson stars as black hole mimickers}",
    eprint = "1009.1250",
    archivePrefix = "arXiv",
    primaryClass = "astro-ph.HE",
    doi = "10.1103/PhysRevD.80.084023",
    journal = "Phys. Rev. D",
    volume = "80",
    pages = "084023",
    year = "2009"
}

@article{Guzman:2005bs,
    author = "Guzman, F. Siddhartha",
    title = "{Accretion disc onto boson stars: A Way to supplant black holes candidates}",
    eprint = "gr-qc/0512081",
    archivePrefix = "arXiv",
    doi = "10.1103/PhysRevD.73.021501",
    journal = "Phys. Rev. D",
    volume = "73",
    pages = "021501",
    year = "2006"
}

@article{Rosa:2022tfv,
    author = "Rosa, Jo{\~a}o Lu{\'\i}s and Rubiera-Garcia, Diego",
    title = "{Shadows of boson and Proca stars with thin accretion disks}",
    eprint = "2204.12949",
    archivePrefix = "arXiv",
    primaryClass = "gr-qc",
    doi = "10.1103/PhysRevD.106.084004",
    journal = "Phys. Rev. D",
    volume = "106",
    number = "8",
    pages = "084004",
    year = "2022"
}

@article{Torres:2002td,
    author = "Torres, Diego F.",
    title = "{Accretion disc onto a static nonbaryonic compact object}",
    eprint = "hep-ph/0201154",
    archivePrefix = "arXiv",
    doi = "10.1016/S0550-3213(02)00038-X",
    journal = "Nucl. Phys. B",
    volume = "626",
    pages = "377--394",
    year = "2002"
}

@article{Herdeiro:2021lwl,
    author = "Herdeiro, Carlos A. R. and Pombo, Alexandre M. and Radu, Eugen and Cunha, Pedro V. P. and Sanchis-Gual, Nicolas",
    title = "{The imitation game: Proca stars that can mimic the Schwarzschild shadow}",
    eprint = "2102.01703",
    archivePrefix = "arXiv",
    primaryClass = "gr-qc",
    doi = "10.1088/1475-7516/2021/04/051",
    journal = "JCAP",
    volume = "04",
    pages = "051",
    year = "2021"
}

@article{Olivares:2018abq,
    author = "Olivares, Hector and Younsi, Ziri and Fromm, Christian M. and De Laurentis, Mariafelicia and Porth, Oliver and Mizuno, Yosuke and Falcke, Heino and Kramer, Michael and Rezzolla, Luciano",
    title = "{How to tell an accreting boson star from a black hole}",
    eprint = "1809.08682",
    archivePrefix = "arXiv",
    primaryClass = "gr-qc",
    doi = "10.1093/mnras/staa1878",
    journal = "Mon. Not. Roy. Astron. Soc.",
    volume = "497",
    number = "1",
    pages = "521--535",
    year = "2020"
}

@article{Torres:2000dw,
    author = "Torres, Diego F. and Capozziello, S. and Lambiase, G.",
    title = "{A Supermassive scalar star at the galactic center?}",
    eprint = "astro-ph/0004064",
    archivePrefix = "arXiv",
    doi = "10.1103/PhysRevD.62.104012",
    journal = "Phys. Rev. D",
    volume = "62",
    pages = "104012",
    year = "2000"
}

@article{Zhang:2025xnl,
    author = "Zhang, Yu-Peng and Wei, Shao-Wen and Liu, Yu-Xiao",
    title = "{Emerging black hole shadow from collapsing boson star}",
    eprint = "2503.14159",
    archivePrefix = "arXiv",
    primaryClass = "gr-qc",
    month = "3",
    year = "2025"
}

@article{Huang:2024wpj,
    author = "Huang, Jiewei and Zhang, Zhenyu and Guo, Minyong and Chen, Bin",
    title = "{Images and flares of geodesic hot spots around a Kerr black hole}",
    eprint = "2402.16293",
    archivePrefix = "arXiv",
    primaryClass = "gr-qc",
    doi = "10.1103/PhysRevD.109.124062",
    journal = "Phys. Rev. D",
    volume = "109",
    number = "12",
    pages = "124062",
    year = "2024"
}

@article{Cardoso:2017njb,
    author = "Cardoso, Vitor and Pani, Paolo",
    title = "{The observational evidence for horizons: from echoes to precision gravitational-wave physics}",
    eprint = "1707.03021",
    archivePrefix = "arXiv",
    primaryClass = "gr-qc",
    month = "7",
    year = "2017"
}

@article{Cunha:2018gql,
    author = "Cunha, Pedro V. P. and Herdeiro, Carlos A. R. and Rodriguez, Maria J.",
    title = "{Does the black hole shadow probe the event horizon geometry?}",
    eprint = "1802.02675",
    archivePrefix = "arXiv",
    primaryClass = "gr-qc",
    doi = "10.1103/PhysRevD.97.084020",
    journal = "Phys. Rev. D",
    volume = "97",
    number = "8",
    pages = "084020",
    year = "2018"
}

@article{Cardoso:2017cqb,
    author = "Cardoso, Vitor and Pani, Paolo",
    title = "{Tests for the existence of black holes through gravitational wave echoes}",
    eprint = "1709.01525",
    archivePrefix = "arXiv",
    primaryClass = "gr-qc",
    doi = "10.1038/s41550-017-0225-y",
    journal = "Nature Astron.",
    volume = "1",
    number = "9",
    pages = "586--591",
    year = "2017"
}

@article{Broderick:2009ph,
    author = "Broderick, Avery E. and Loeb, Abraham and Narayan, Ramesh",
    title = "{The Event Horizon of Sagittarius A*}",
    eprint = "0903.1105",
    archivePrefix = "arXiv",
    primaryClass = "astro-ph.HE",
    doi = "10.1088/0004-637X/701/2/1357",
    journal = "Astrophys. J.",
    volume = "701",
    pages = "1357--1366",
    year = "2009"
}

@article{Zhang:2021xhp,
    author = "Zhang, Yu-Peng and Zeng, Yan-Bo and Wang, Yong-Qiang and Wei, Shao-Wen and Liu, Yu-Xiao",
    title = "{Motion of test particle in rotating boson star}",
    eprint = "2107.04848",
    archivePrefix = "arXiv",
    primaryClass = "gr-qc",
    doi = "10.1103/PhysRevD.105.044021",
    journal = "Phys. Rev. D",
    volume = "105",
    number = "4",
    pages = "044021",
    year = "2022"
}

\end{document}